\documentclass[usenatbib,onecolumn]{mnras}

\usepackage[T1]{fontenc}
\usepackage{ae,aecompl}

\usepackage{amsmath}
\usepackage{amssymb}
\usepackage{braket}
\usepackage{graphicx}
\usepackage{booktabs}
\usepackage{multicol}
\usepackage{hyperref}
\usepackage{bm}
\usepackage{color}
\usepackage{mathrsfs}
\usepackage{subcaption}
\newcommand\be{\begin{equation}}
\newcommand\ee{\end{equation}}
\newcommand\mrm{\mathrm}
\newcommand\mbf{\mathbf}
\newcommand\dif{\mrm{d}}

\newcommand{\mat}[1]{{\mathbf{#1}}}
\newcommand\hz{\hat{\bm{z}}}

\newcommand\dA{{\bm{d}}_A}

\newcommand\n{\hat{\bm{n}}}
\newcommand\calC{\vartheta}

\newcommand\calN{\mathcal{N}}
\newcommand\calS{\mathcal{S}}
\newcommand\Sig{\mathbf\Sigma}
\newcommand\Sigyy{\mathbf{\Sigma_{yy}}}
\newcommand\Sigrr{\mathbf{\Sigma_{rr}}}
\newcommand\Sigrri{\mathbf{\Sigma}_{\mathbf{rr}}^{-1}}
\newcommand\Sigyyi{\mathbf{\Sigma}_{\mathbf{yy}}^{-1}}
\newcommand\Crr{\mathbf{C_{rr}}}
\newcommand\Cyy{\mathbf{C_{yy}}}
\newcommand\Cry{\mathbf{M}}

\newcommand\Cyyi{\mathbf{C}_{\mathbf{yy}}^{-1}}
\newcommand\CryT{\mathbf{M}^\T}
\newcommand\Eyy{\mathbf{E_{yy}}}
\newcommand\Eyyi{\mathbf{E}_{\mathbf{yy}}^{-1}}

\newcommand\cov{\mathbf{C}}

\newcommand\R{\mathbf{R}}
\newcommand\A{\mathbf{A}}
\newcommand\M{\mathbf{W}}
\newcommand\E{\mathbf{E}}
\newcommand\V{\bm{V}}
\newcommand\U{\bm{U}}
\newcommand\T{\mathsf{T}}

\newcommand\x{\mathbf{x}}

\newcommand\z{\bm{z}}

\newcommand\hn{\hat{\bm{n}}}
\newcommand\hx{\hat{\mbf{x}}}

\newcommand\hth{\hat{\theta}}
\newcommand\thhat{\bm{\hat{\theta}}}
\newcommand\rhat{\bm\hat{r}}
\newcommand\shat{\bm{\hat{s}}}
\newcommand\ihat{\bm{\hat{\imath}}}
\newcommand\zhat{\bm{\hat{z}}}
\newcommand\rbar{\bm{\bar{r}}}
\newcommand\sbar{\bm{\bar{s}}}
\newcommand\ibar{\bm{\bar{\imath}}}
\newcommand\zbar{\bm{\bar{z}}}
\newcommand\xbar{\mathbf{\bar{x}}}
\newcommand\wbar{\bm{\bar{w}}}

\newcommand\DelO{\bm{\Delta_0}}
\newcommand\logl{L}

\newcommand\hs{\hat{s}}
\newcommand\hi{\hat{\imath}}
\newcommand\br{\bar{r}}
\newcommand\bs{\bar{s}}
\newcommand\bi{\bar{\imath}}
\newcommand\bmi{\bm{i}}
\newcommand\s{\bm{s}}

\newcommand\ybar{\mathbf{\bar y}}
\newcommand\yhat{\mathbf{\hat y}}

\newcommand\mutilde{\bm{\tilde\mu}}

\def\y{\mathbf{y}}
\def\v{\bm{v}}
\def\w{\bm{w}}
\def\k{\bm{k}}

\def\r{\bm{r}}

\def\Pr{p}

\newcommand\LCDM{$\Lambda$CDM}
\newcommand\HO{H_0}

\title[Cosmological inference of peculiar velocities]{A probabilistic framework for
cosmological inference of peculiar velocities}

\author[L.~Dam]{Lawrence Dam$^{1}$\thanks{E-mail: ldam4036@uni.sydney.edu.au}
\\
$^{1}$Sydney Institute for Astronomy, School of Physics, A28,
			The University of Sydney, NSW 2006, Australia\\
}

\date{Accepted XXX. Received YYY; in original form ZZZ}

\pubyear{2020}

\begin{document}
\label{firstpage}
\pagerange{\pageref{firstpage}--\pageref{lastpage}}
\maketitle

\begin{abstract}
We present a Bayesian hierarchical framework for a principled data analysis pipeline
of peculiar velocity surveys, which makes explicit the inference problem of
constraining cosmological parameters from redshift-independent distance indicators. We demonstrate our
method for a Fundamental Plane-based survey. The essence of our approach is to 
work closely with observables (e.g.\ angular size, surface brightness, redshift, etc),
through which we bypass the use of summary statistics by working with the probability
distributions. The hierarchical approach improves upon the usual analysis
in several ways. In particular, it allows a consistent analysis without
having to make prior assumptions about cosmology during the calibration phase.
Moreover, calibration uncertainties are correctly accounted for in
parameter estimation. Results are presented for a new, fully analytic
posterior marginalised over all latent variables, which we expect to allow
for more principled analyses in upcoming surveys. A maximum a posteriori estimator 
is also given for peculiar velocities derived from Fundamental Plane data.
\end{abstract}


\section{Introduction}

The gravitational pull of large-scale structure perturbs the motion of galaxies
away from the Hubble flow giving rise to so-called peculiar velocities.
The presence of peculiar velocities complicates the recovery of distances to
galaxies, type Ia supernovae and other objects, but in themselves can be exploited
as unbiased cosmological probes of the underlying total matter density field.
Some of the ways peculiar velocities have been used include: measuring the
growth rate of structure from the velocity power spectrum
\citep{Koda:2013eya,Johnson:2014kaa,Howlett:2017} and the density-weighted
velocity power spectrum \citep{Qin:2019};
comparisons between predictions from the density field and observed velocity field
(\cite{Willick:1996km,Willick:1998}; see also the review by
\cite{Strauss:1995fz}, and references therein); cosmological constraints
from observed versus predicted velocity comparisons \citep{Carrick:2015} and velocity-density
cross-correlations \citep{Nusser:2017,Adams:2017val}; reconstruction of the
local velocity field
\citep{Fisher:1995,Zaroubi:1999,Dekel:1999,Courtois:2013,Hoffman:2017ako};
testing statistical isotropy \citep{Schwarz:2007wf,Appleby:2014kea,Soltis:2019ryf};
testing modified gravity \citep{Hellwing:2014,Johnson:2015aaa};
consistency tests of \LCDM\ \citep{Nusser:2011tu,Huterer:2016uyq}.

Peculiar velocities simultaneously affect both observed redshift and the inferred
distance through the Doppler effect and relativistic beaming effect. (There is
also a relativistic aberration effect caused by the observer's own peculiar motion,
which induces a lensing-like deflection; this is straightforward to account for but
will be unimportant in this paper so will be ignored.)
The problem is to separate out the contribution from the peculiar velocity $v$ from
the unobserved cosmological, background contribution, in the
presence of measurement uncertainties and systematics.
The observed redshift is given by $1+z\simeq(1+\bar{z})(1+v/c)$,
where $\bar{z}$ is the cosmological redshift, and for typical objects $v/c\sim10^{-3}$.
Together with independent knowledge of the distance, which fixes the distance-redshift
relation, we can invert to obtain $v$.

Departures away from homogeneity in the Universe source fluctuations in the distance
to galaxies. Besides peculiar velocities there are also contributions from
gravitational lensing, gravitational redshift,
the Sachs-Wolfe effect and an integrated Sachs-Wolfe-like effect
\citep{Sasaki:1987ad,Sugiura:1999a,Hui:2005nm}. At low redshifts ($z\lesssim0.1$)
the dominant contribution is from peculiar velocity, while at high redshifts ($z\gtrsim1$)
it is from gravitational lensing \citep{Bolejko:2012uj}. While the change to the observed
redshift is small ($\sim10^{-3}$), the change to distance is typically a few percent
and, for a \LCDM\ cosmology, can be up to $10\%$. However, it is at these low-redshifts
that peculiar velocities as cosmological probes are perfectly suited because:
(\emph{i}) velocities are sourced directly from the total matter fluctuations
so are expected to be unbiased tracers of the mass distribution \citep{Peebles:1993};
(\emph{ii}) distance fluctuations due to peculiar motion are dominant at low redshifts
\citep{Bacon:2014uja}; (\emph{iii}) distance measurement uncertainties grow
with redshift \citep{Scrimgeour:2015khj}; and
(\emph{iv}) the velocity correlation function is more sensitive to large scales than
its density counterpart since in Fourier space $v\sim \delta/k$. These factors mean
that the signal-to-noise ratio -- with
signal being the fluctuations to distance caused by peculiar velocities -- tends to
increase with decreasing redshift as $\sim 1/z$. As such, peculiar velocity catalogues do
not necessarily benefit from a larger survey depth but, nevertheless, are excellent
probes of cosmology based on observations of the low redshift Universe.

Conventionally, peculiar velocities have been estimated as the residual motion of
Hubble's law,
\be
v \simeq cz - {\HO}d .
\ee
The appearance of Hubble's constant $\HO$ besides the distance $d$ means
that peculiar velocities are independent of the absolute calibration of distances.
Moreover this relation contains several approximations, as pointed out by \cite{Davis:2014jwa}
and elsewhere, and is of limited accuracy for precision cosmology with current and
future surveys. However, here it serves to illustrate that typically peculiar velocities are
estimated roughly as some difference between the total (observed) and the cosmological
background (not observed). These velocities are not random but are coherently sourced
from the underlying matter density field meaning that nearby objects will have
correlated $v$. As we show, this is one idea that can be exploited when jointly
estimating the velocities $v_1,v_2,\ldots, v_N$ from a sample of $N$ objects.

\subsection{Motivation}

A recent trend in cosmology is towards performing principled Bayesian
inference. Some recent examples include cosmological analysis of type Ia supernovae
\citep{Mandel:2009,March:2011,Sharif:2016,Hinton:2019}, cosmic shear
\citep{Schneider:2015,Alsing:2016}, large-scale structure
\citep{Jasche:2010,Jasche:2013}; estimating $\HO$ from the cosmic distance
ladder \citep{Feeney:2017sgx}; estimating photometric redshifts and redshift
distribution \citep{Leistedt:2016,Sanchez:2019}. Such approaches will be
important for maximising the scientific return of upcoming surveys and
ensuring that conclusions are robust, particularly for blinded analyses
and tests of the \LCDM\ model. It is also important that the calibration, validation,
and ultimately the production of catalogues makes a minimal amount of model-dependent
assumptions and that uncertanties related to this phase are correctly propagated.
Model comparison and parameter estimation are two different tasks and it is
desirable when testing between competing models that the data does not contain
implicit assumptions that could bias inference.

The method we describe here is developed for cosmological analysis based on
peculiar velocities with the foregoing concerns in mind. To build a catalogue
of peculiar velocities requires a measure of the distance to the source, and
this can be obtained from  type Ia supernovae or redshift-independent distance
indicators, which relate the luminosity (as in the case of the Tully-Fisher
relation) or the size (as in the case of the Fundamental Plane relation), to
other intrinsic, physical properties of galaxies.
A probabilistic framework for the Tully-Fisher relation has been developed
in the past \citep{Willick:1994}, and clarifies the problem of
calibrating the relation and then using it to estimate distances. Aspects of this
important early approach are similar to Bayesian hierarchical models, which
have steadily gained in prominence in cosmology in recent years (see, e.g.\
\cite{Loredo:2012} and references therein).

In this work we focus on the Fundamental Plane relation. While it is common
to refer to this relation as a distance indicator it is more accurately
termed a size indicator \citep{Hobson:2010}. 
The size is a physical characteristic of the source object and in itself
does not depend on redshift. To fit the relation, however, does require redshift
and a model prior to convert to distances; multiplying the distance by
the observed angular size then gives the physical size.

While the dependence of distance on cosmology is generally weak at the
redshifts concerned, in seeking a more principled approach it is clearly more
desirable to allow the data to determine the best cosmological model in the
first place. Because there are various scientific goals besides cosmology, here
we will demonstrate one method for how to directly use the Fundamental Plane data
to perform inference in
a way that allows a joint fit to the Fundamental plane and the cosmological model,
which would otherwise be assumed during calibration phase. In principle, if
the aim is to study the properties of galaxies, in which case the peculiar
velocities are a nuisance, then one simply marginalises over the cosmological
parameters, whereas if the aim is to use peculiar velocities as a probe
then one marginalises over the Fundamental Plane parameters.
This idea of jointly calibrating and fitting the cosmological model is
not new. For example, the recovery of the CMB power spectrum requires nuisance
parameters modelling calibration, beam uncertainties, foreground
power spectrum templates, and these are optimised at the same time as
the cosmological parameters \citep{Likelihood:2015}; a more recent example
is the estimation of $\HO$ from a global fit of the cosmic
distance ladder, which involves calibrating the Cepheid Leavitt law and
the type Ia supernovae Tripp relation \citep{Zhang:2017,Feeney:2017sgx}.

The advantage of this approach is that, because the calibration is not absolute,
uncertainties in the Fundamental Plane data can be carried downstream,
allowing us, at least in principle, to build a probabilistic catalogue of
the peculiar velocities \citep{Brewer:2012gt,Portillo:2017}. Although a
peculiar velocity catalogue delivers posterior information (i.e.\ contains
a model prior), in our approach model assumptions are made transparent,
allowing different cosmological models to be more robustly tested.

Although our approach is presented in the case of the Fundamental Plane relation,
we expect that methods described here can be applied without significant modifications
to the Tully-Fisher relation, as well. In both cases, calibration of each relation
amounts to fitting a linear relation, but the difference for the Tully-Fisher
relation is that the data is univariate and the fit is to a line rather than a
plane.

The rest of this paper is organized as follows.
In Section \ref{sec:PV-FP} we give a brief review of the Fundamental
Plane relation and its calibration using the maximum likelihood method. In
Section \ref{sec:bhm} we develop a framework for constraining cosmology
from the Fundamental Plane. For the reader not interested in details of the
calculation, the main result \eqref{eq:post} in this section is a new
posterior for performing joint fits. (Note in this section we assume the model is a \LCDM\
cosmology, but we emphasize that because the conditional data does not have
model assumptions built in, any non-\LCDM\ model can also be used in this
framework, by modifying to the appropriate distance-redshift relation and
prior for peculiar velocity statistics.)
In Section \ref{sec:expts} we present some numerical results for a mock
analysis. In Section \ref{sec:selection} we discuss how the framework can be
generalised to include selection effects that Fundamental Plane data are
affected by. In Section \ref{sec:conclusions} we conclude and summarise our main results.

Throughout this paper we work with a spatially flat \LCDM\ cosmology ($\Omega_k=0$)
for simplicity and take the observer's peculiar velocity $\v_O$ to be zero so that
quantities are as measured in the idealised cosmic rest frame (`CMB frame').
\newline
\newline
\textit{Notation.}
\noindent A source observed in the direction $\hn$ has a line-of-sight peculiar
velocity denoted $v=\v\cdot\hn$. Sources (e.g.\ galaxies) are labelled by
subscript $m,n,\ldots$, while $i,j,\ldots$ are reserved for the components of 
spatial vectors; e.g.\ $v^i_m$ denotes the
$i^\mrm{th}$ component of the $m^\mrm{th}$ source's velocity $\v_m$.
Unless otherwise specified $r$ denotes the logarithm of the effective
radius of galaxies, and we use $c$ for both the zero-point of the
Fundamental Plane relation and the speed of light, although it will be clear from the
context which is being used. Vectors are typeset using boldface, while
matrices are typeset $\mat{A},\Sig,\mat{C},\ldots$, and are always
denoted by uppercase symbols. For convenience, in Table \ref{tab:symbols}
we provide a summary of notation used in this work.

\section{The Fundamental Plane}\label{sec:PV-FP}

The Fundamental Plane (FP) relation is an observed correlation
of elliptical galaxies between its effective size $R_e$ defined such
that it contains half of the total galaxy luminosity (`half-light'),
the central velocity dispersion $\sigma_0$, and the mean surface brightness 
$\langle I_e\rangle$ enclosed within the effective radius. That such an
empirical correlation exists might be expected from the virial theorem,
\be
\sigma^2_\mrm{vir}
\propto \frac{GM}{2R}
\propto R \left(\frac{M}{L}\right) \left(\frac{L}{R^2}\right),
\ee
provided that the mass-to-light ratio $M/L$ is constant, and that both the virial size
$R$ and velocity dispersion $\sigma_\mrm{vir}$ are proportional to $R_e$ and $\sigma_0$,
respectively. In terms of logarithmic quantities, the FP relation is 
\citep{Dressler:1987,Djorgovski:1987}
\be\label{eq:FP}
r=as+bi+c,
\ee
where $r\equiv\log R_e$, $s\equiv\log\sigma_0$, and $i\equiv\log\langle I_e\rangle$.
The coefficients $a$ and $b$ define the orientation of the plane in
$(r,s,i)$-space, and $c$ determines the height (the so-called zero-point).
The distance-independent observables are $s$ and $i$ and can be directly measured;
the logarithmic physical size $r$ of the galaxy, however, is to be inferred
from \eqref{eq:FP} once it has been calibrated. What is actually observable
is the \emph{angular} size $\theta$ of the galaxy, and this is related to $r$ through
the angular diameter distance:
\be\label{eq:r}
r=\log\theta + \log d_A.
\ee
Thus the FP relation can be used as a distance indicator. On top of the
intrinsic scatter already present, peculiar velocities induce an additional
source of scatter to the FP, which can be used to infer the peculiar velocity.

The problem of deriving peculiar velocities from FP data is that \eqref{eq:FP}
is a size indicator, and the conversion to distance cannot be done without
first assuming a cosmological model. As a model prior is already built
into the calibration the nominal catalogue data might be thought to
deliver posterior information, rather than likelihood information.
The calibration of the FP relation \eqref{eq:FP} typically makes an
assumption about cosmological model in order to convert observed angular
size to physical size $r$ \citep{Magoulas:2012jy,Springob:2014qja}.
While the cosmological dependence of distance at low redshifts
may be weak, this ignores the statistical fluctuations of the peculiar
velocities, which is sensitive to cosmology through the power spectrum, i.e.\
velocities are not randomly sourced.
Our method, which we present in Section~\ref{sec:bhm}, improves upon
the standard approach by making no assumption about cosmological model
during this conversion step through making use of data further
upstream, namely, the velocity dispersion, surface brightness, angular
size, and angular coordinates.
As our approach takes as basic input observables directly related to the FP
relation, we will first review how the FP is typically used to estimate
peculiar velocities.

\subsection{Fundamental Plane maximum likelihood method}
The calibration of the FP will be based on a well-tested maximum likelihood (ML) method
that was developed in \cite{Saglia:2000fa} and used by \cite{Colless:2000et}, and
more recently by the 6-degree Field Galaxy Survey (6dFGS) \citep{Springob:2014qja}. The
likelihood of obtaining the data $(\hat{r}_m,\hs_m,\hi_m)$ for the $m^\mrm{th}$
object is given by a truncated trivariate Gaussian
\be\label{eq:FP-ML}
\Pr(\hx_m)
=\frac{1}{(2\pi)^{N/2}\det(\cov^\mrm{FP}+\E^\mrm{FP}_m)^{1/2}}
\exp\left[
		-\frac{1}{2}(\hx_m-\xbar)^\T(\cov^\mrm{FP}+\E^\mrm{FP}_m)^{-1}(\hx_m-\xbar)
	\right]\,
f_m^{-1}\prod_{j=1}^M \Theta(\mbf{u}_j^\T\hx_m-w_j),
\ee
where $\hx_m=(\hat{r}_m,\hs_m,\hi_m)^\T$, $\xbar=(\br,\bs,\bi)^\T$,
$\cov^\mrm{FP}$ describes the FP and its intrinsic scatter, and $\E^\mrm{FP}_m$
gives the measurement errors; the presence of the Heaviside step function
$\Theta(\cdots)$ is to enforce the selection criteria, because of which
$f_m$ is needed to ensure that $\int\Pr(\hx_m)\,\dif^3\,\hx_m=1$.
There are two (linear) constraints on the observable part of the FP that are
usually considered; they are due to a cutoff in the measurable velocity
dispersion and magnitude. Other selection criteria may also be included
depending on the instrumental setup.
Here the vector $\mbf{u}_j$ depends on the FP parameters. The likelihood
\eqref{eq:FP-ML} can be understood as the convolution between the Gaussian
error distribution and the Gaussian population distribution,
with centroid $(\bar{r},\bs,\bi)$ and covariance of $\cov^\mrm{FP}$. In the language
of hierarchical modeling \citep{Loredo:2004,Hogg:2010ma} this is the probability
of obtaining the data after marginalising over the true variables,
with the parameters $\br,\bs,\bi,\sigma_1,\sigma_2$, and $\sigma_3$
considered hyperparameters. The use of \eqref{eq:FP-ML} is motivated
by the fact that the data $\{r,s,i\}$ appears to be Gaussian distributed to
a good approximation \citep{Colless:2000et}.
The ML method essentially fits a 3-dimensional ellipsoid to the data,
with the centroid corresponding to the mean of the Gaussian and the
principal axes aligned with the eigenvectors of the covariance matrix.

In general, because the FP is tilted with respect to the FP-space axes
defined by $r$, $s$ and $i$, the covariance matrix will contain off-diagonal
entries that are functions of the orientation parameters $a$ and $b$.
The intrinsic scatter is relative to the two axes spanning the plane
and the axis normal to the plane.\footnote{There is some arbitrariness
in how one chooses the vectors that span the FP \citep{Saglia:2000fa} but
does not affect the fit to the FP coefficients.}
The trivariate Gaussian may be diagonalised by a rotation of the data
${\hx_m\to\hx'_m=\mat{O}^\T\hx_m}$,
such that in these new coordinates the covariance is diagonalised
$\cov^\mrm{FP}\to\mat{D}=\mat{O}^\T\cov^\mrm{FP}\mat{O}
=\mrm{diag}(\sigma_1^2,\sigma_2^2,\sigma_3^2)$,
with $a$ and $b$ partially describing the rotation matrix $\mat{O}$;
more details can be found in Appendix~\ref{app:geometry}.
In this approach it can be seen that $a$ and $b$ are related to the
correlation coefficients, and that constraining the population distribution
simultaneously fits the FP as a by-product. (Note that if $a=0$, $b=0$ then $\cov^\mrm{FP}$
is diagonal and there exists no statistical relation between $r,s,i$.)
Thus there are eight parameters to be determined: $(\bar{r},\bs,\bi)$
specifies the centroid, and $(\sigma_1,\sigma_2,\sigma_3,a,b)$
determines the intrinsic scatter and orientation of the FP.
Since the FP relation provides the constraint $\br=a\bs+b\bi+c$, the zero-point
$c$ is a derived parameter, i.e.\ we can either parametrize using $\br$ or $c$,
though the covariance of $c$ with $a$ is strong and will be less well behaved
in sampling space. (Below we will use $\br$, but if instead we use $c$ then
$\br$ should be replaced by $a\bs+b\bi+c$.)
Note that since $\cov^\mrm{FP}$ is a $3\times3$ symmetric matrix it is
specified by six parameters but here we have only five parameters;
the additional parameter is due to the rotational degree of freedom
allowing rotations of the (infinite) plane on to itself.\footnote{%
The rotation matrix $\mat{O}$ can be written as the composition of
three rotation matrices,
each specified by an Euler angle. As only two of these angles are
constrained we are free to fix the third.}

These parameters are obtained by maximising the (logarithm of the)
joint likelihood of all $N$ objects in the sample:
\be\label{eq:lnL-FP}
\ln\mathcal{L}=\sum_{m=1}^N \ln \Pr(\hx_m).
\ee
This calibration step is performed in the process of building the
catalogue, and assumes some fiducial cosmological model. Since the goal is to
estimate cosmological parameters, a principled approach is to thus
jointly calibrate the FP relation and perform the analysis simultaneously
-- similar to how the zero-point of standard candles (absolute magnitude)
are constrained along with cosmological parameters. This idea of a global
fit (i.e.\ not separating calibration from parameter estimation) is therefore
not new, but has yet to be applied to peculiar velocity cosmology to our
knowledge. It is the goal of this work to address this problem.

Already the likelihood \eqref{eq:FP-ML} suggests the use of a hierarchical
approach: the FP parameters $a$ and $b$ are to be estimated along with
population hyperparameters. As we mentioned above, the likelihood in
\eqref{eq:lnL-FP} can be viewed as the marginalisation over latent
variables:
\be\label{eq:FP-conv}
\mathcal{L} = \prod_{m=1}^N \int \dif^3\x_m \:
			 f(\x_m\mid a,b,\br,\bs,\bi,\sigma_1,\sigma_2,\sigma_3)\,
			 \ell(\hx_m\mid \x_m),
\ee
where $f$ is the population-level distribution and the individual
likelihoods $\ell(\hx_m\mid \x_m)$ gives the error distribution.

The 6dFGS and upcoming Taipan survey \citep{daCunha:2017wwy} will derive
peculiar velocity estimates from the FP relation. The observables are
the PDF of the ratio of the observed effective radius
$R$ to the inferred physical radius $\bar{R}$. It is shown that
the \emph{logarithmic} ratio $\log(R/\bar{R})$ is Gaussian
distributed and not of $R/\bar{R}$. This is then related to the PDF
of the cosmological distance ratio
\be\label{eq:eta}
\hat\eta\equiv\log D(\hat{z})/D(\bar{z}),
\ee
where $D$ is the comoving distance. The probabilistic outputs of the
6dFGS catalogue are not of the peculiar velocities (which are
significantly skewed) but of $\hat\eta$ (well-described by a Gaussian).
Nevertheless, there is a slight skew but small enough that modelling the
PDF as a Gaussian should be adequate \citep{Springob:2014qja}.
For each source the PDF is summarised by the mean and standard
deviation of a Gaussian, and also higher-order moment
given by the skew parameter $\alpha$ of the Gram-Charlier series.%
Summary statistics used in catalogues are valid provided the underlying
PDF has (approximately) Gaussian uncertainties; if this is not the case
then one may still be able to find a transformation of the data so that it is.

Since the effective size $R_e$ is physical the effect of an object's peculiar
velocity is to modify its angular size $\theta$ and
inferred distance $d_A$: there is no change in $r$. Since the fractional
changes are of equal size but opposite sign they
cancel to first-order so that $r$ is constant. For an enlightening
discussion we refer the reader to \cite{Kaiser:2014jca}.

\begin{table}
\caption{A summary of commonly used mathematical symbols.
        In the third column we give the equation in which the symbol is first
        used or otherwise defined nearby. \label{tab:symbols}}
\begin{tabular*}{\textwidth}{l @{\extracolsep{\fill}} c c}
\hline\hline
$a$, $b$, $c$ & Coefficients of the FP relation & \eqref{eq:FP} \\
$\sigma_1$, $\sigma_2$, $\sigma_3$ & Intrinsic standard deviations of the
trivariate Gaussian & \eqref{eq:FP-ML} \\
$\xbar=(\br,\bs,\bi)^\T$ & Centroid of FP population distribution & \eqref{eq:FP-ML} \\
$\mathcal{C}$ & Set of cosmological parameters & \eqref{eq:pars} \\
$\mathcal{F}$ & Set of parameters related to the FP & \eqref{eq:pars} \\
$\vartheta$ & Set of FP-related parameters and cosmological parameters & \eqref{eq:pars} \\
$z$ ($\hat{z}$) & Latent (observed) total redshift \\
$\bar{z}$ & Background (i.e.\ cosmological) redshift \\
$\theta$ ($\hat{\theta}$) & Latent (observed) angular size & \eqref{eq:r} \\
$r$ ($\rhat$) & Latent (observed) logarithm of the effective (half-light) radius & \eqref{eq:FP} \\
$s$ ($\hs$) & Latent (observed) logarithm of the velocity dispersion $\sigma_0$ & \eqref{eq:FP} \\
$i$ ($\hi$)& Latent (observed) mean surface brightness & \eqref{eq:FP} \\
$\alpha$ ($\hat\alpha$)& Latent (observed) right ascension \\
$\delta$ ($\hat\delta$)& Latent (observed) declination \\
$\cov^\mrm{FP}$ & The $3\times3$ covariance matrix of $r$, $s$ and $i$ & \eqref{eq:FP-ML} \\
$\E^\mrm{FP}_m$ & The $3\times3$ covariance matrix of experimental errors of
$\rhat$, $\hs$ and $\hi$ of the
    $m^\mrm{th}$ galaxy & \eqref{eq:post} \\
$\cov$ & The $2\times2$ covariance matrix of $s$ and $i$; submatrix of $\cov^\mrm{FP}$ & \eqref{eq:post} \\
$\E_m$ & The $2\times2$ covariance matrix of experimental errors of $\hs$ and $\hi$ of the
    $m^\mrm{th}$ galaxy; submatrix of $\E^\mrm{FP}_m$ & \eqref{eq:post} \\
$\mathscr{D}$ & Observed data \\
$\s$ & The $N$-dimensional column vector $(s_1,s_2,\ldots,s_N)^\T$ & \eqref{eq:post-unmarg} \\
$\bmi$ & The $N$-dimensional column vector $(i_1,i_2,\ldots,i_N)^\T$ & \eqref{eq:post-unmarg} \\
$\rbar$ & The $N$-dimensional column vector $(\br,\br,\ldots,\br)^\T$ \\
$\sbar$ & The $N$-dimensional column vector $(\bs,\bs,\ldots,\bs)^\T$ \\
$\ibar$ & The $N$-dimensional column vector $(\bi,\bi,\ldots,\bi)^\T$ \\
$\y$ & The $2N$-dimensional column vector $(\s,\bmi)^\T$ & \eqref{eq:shifted-r-moments} \\
$\ybar$ & The $2N$-dimensional column vector $(\sbar,\ibar)^\T$ & \eqref{eq:shifted-r-moments} \\
$\Crr$ & The $N\times N$ covariance matrix of $\r$ & \eqref{eq:Co} \\
$\Cyy$ & The $2N\times 2N$ covariance matrix of $\s$ and $\bmi$ & \eqref{eq:Co} \\
$\Cry$ & The $2N\times N$ matrix of covariances of $\r$ and $(\s,\bmi)$ & \eqref{eq:Co} \\
$\Eyy$ & The $2N\times 2N$ covariance matrix of experimental errors of $\shat$ and $\ihat$ & \eqref{eq:p-si-hatted} \\
$\R$ & The $N\times N$ covariance matrix of peculiar velocities $\V$ & \eqref{eq:pec-vel-like} \\
$\bm{\logl_\theta}$ & The $N$-dimensional vector
    $\big(\log\hth_1, \log\hth_2, \ldots, \log\hth_N\big)^\T$ & \eqref{eq:log_theta} \\
$\bm{\logl_d}$ & The $N$-dimensional vector
    $\big(\log{d}_A(\hat{z}_1),\log{d}_A(\hat{z}_2),\ldots,\log{d}_A(\hat{z}_N)\big)^\T$ & \eqref{eq:log_dA} \\
$\bm{\logl_{\bar{d}}}$ & The $N$-dimensional vector
    $\big(\log\bar{d}_A(\hat{z}_1),\log\bar{d}_A(\hat{z}_2),\ldots,\log\bar{d}_A(\hat{z}_N)\big)^\T$ & \eqref{eq:l-dA} \\
$D(z)$ & Comoving distance to redshift $z$ & \eqref{eq:eta} \\
$d_A$ & Total angular diameter distance \\
$\bar{d}_A$ & Background (i.e.\ unperturbed) angular diameter distance \\
$\hn$ & Unit direction vector in $\mathbb{R}^3$ \\
$\bm{x}$ & Comoving position vector \\
$\v$ & Peculiar velocity vector \\
$v=\hn\cdot\v$ & Line-of-sight peculiar velocity \\
$\V$ & The $N$-dimensional column vector $(v_1,v_2,\ldots,v_N)^\T$ & \eqref{eq:post-unmarg} \\
$\x=(r,s,i)^\T$ & Vector of latent FP observables & \eqref{eq:FP-conv} \\
$\hx=(\rhat,\hs,\hi)^\T$ & Vector of observed FP observables & \eqref{eq:FP-ML} \\
$\calN(\x\,;\,\bm\mu,\Sig)$ & Multivariate Gaussian probability density function with
    mean $\bm\mu$ and covariance $\Sig$ \\
$\Theta(\cdot)$ & The Heaviside step function & \eqref{eq:FP-ML} \\
$\delta_D(\x)$ & The $N$-dimensional Dirac delta function with $\x$ an $N$-dimensional vector \\
\hline\hline
\end{tabular*}
\end{table}

\section{Cosmological inference directly from the Fundamental Plane}\label{sec:bhm}
To infer cosmological parameters from the FP typically requires a catalogue of
peculiar velocities. When the goal is to constrain the cosmological model
peculiar velocities are summary statistics derived from the FP.
In this section we derive a joint posterior distribution for the
parameters $\vartheta$ (both cosmological and FP) that bypasses the need
for a catalogue. The main result is \eqref{eq:post}. Here each source object
has a peculiar velocity that is treated
as an unknown parameter, which can be marginalised over in the final inference, as
we will show. These can, however, be left explicit at the cost of dealing with
a high-dimensional parameter space.

\begin{figure}
   \centering
   \includegraphics[scale=1.5]{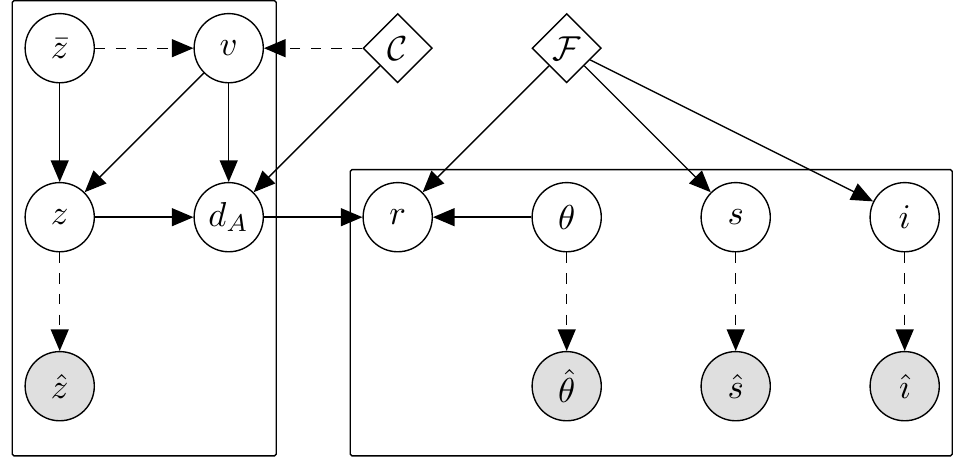}
   \caption{Graphical network showing the structure of the statistical model.
   Dashed lines show probabilistic relations, while
   solid lines show deterministic relations, between observed variables
   (shaded nodes) and latent variables (unshaded nodes).
   To illustrate the parameter dependence we have distinguished
   between cosmological parameters $\mathcal{C}$ and parameters related to
   the FP, $\mathcal{F}=\{a,b,\br,\bs,\bi,\sigma_1,\sigma_2,\sigma_3\}$.
   Note that because we are working to first-order precision
   we can bypass the background redshift $\bar{z}$ altogether by evaluating
   $v$ at the total redshift $z$.}
   \label{fig:FP}
\end{figure}

Suppose we have a sample of $N$ objects with the following data:
\begin{enumerate}
\item the measured redshifts
$\bm\zhat=(\hat{z}_1,\hat{z}_2,\ldots,\hat{z}_N)^\T$;
\item the observed angular sizes
$\thhat=(\hth_1,\hth_2,\ldots,\hth_N)^\T$;
\item the (logarithm of the) velocity dispersions
$\shat=(\hs_1,\hs_2,\ldots,\hs_N)^\T$;
\item the (logarithm of the) surface brightnesses
$\ihat=(\hat\imath_1,\hat\imath_2,\ldots,\hat\imath_N)^\T$;
\item the angular positions $(\alpha_m,\delta_m)$ for each galaxy $m$.
\end{enumerate}
We assume the angular positions $\{\alpha_m,\delta_m\}$ of the objects
are precisely known, treating it as prior information. Thus we seek an
expression for the posterior probability of parameters given the
data:
\be\label{eq:FP-post-CV}
\Pr(\calC\mid\bm\zhat,\thhat,\shat,\ihat).
\ee
The most straightforward way to derive an expression for \eqref{eq:FP-post-CV}
is to begin with the unmarginalised joint posterior
\be\label{eq:post-unmarg}
\Pr(\calC,\V,\r,\s,\bmi,\bm\theta,\dA,\z \mid \zhat,\thhat,\shat,\ihat)
\ee
and use the chain rule to decompose into simpler terms. All unobserved
variables, including the set of line-of-sight velocities
$\V=(v_1,v_2,\ldots,v_N)^\T$ will be marginalised over.
The dependencies
between variables are shown in Fig.~\ref{fig:FP}. For brevity $\calC$
collects all parameters, including the cosmological parameters
$\mathcal{C}=\{\Omega_m,n_s,\sigma_8,\ldots\}$,
the FP relation parameters $\mathcal{F}$, which consist of $a,b$ and the
hyperparameters of the intrinsic population distribution
$\br,\bs,\bi,\sigma_1,\sigma_2,\sigma_3$:
\be\label{eq:pars}
\calC=\{\Omega_m,n_s,\sigma_8,\ldots,a,b,\br,\bs,\bi,\sigma_1,\sigma_2,\sigma_3\}.
\ee
In this work we will be interested in the cosmological parameters but
the calibration parameters are treated on an equal footing and we do not
marginalise over them.

In the following we derive an analytic expression for the
joint posterior for parameters, ignoring for now selection effects;
the inclusion of selection effects is discussed in Section \ref{sec:selection}.
We begin by applying Bayes' theorem to \eqref{eq:post-unmarg}:
\be\label{eq:P-joint-all}
\Pr(\calC,\V,\r,\s,\bmi,\bm\theta,\dA,\z\mid{\bm\zhat},\bm\hth,\shat,\ihat)
\propto
\Pr(\hz,\thhat,\shat,\ihat \mid \calC,\V,\r,\s,\bmi,\bm\theta,\dA,\z) \,
	\Pr(\calC,\V,\r,\s,\bmi,\bm\theta,\dA,\z).
\ee
The first term on the RHS of \eqref{eq:P-joint-all} is related to the observed
data and its experimental errors. Since redshift and angular size measurements
are separable from the other observables we can write
\be\label{eq:P-FP-exp}
\Pr(\hz,\thhat,\shat,\ihat \mid \calC,\V,\r,\s,\bmi,\bm\theta,\dA,\z)
= \Pr(\zhat\mid\z)\, \Pr(\thhat\mid\bm\theta) \, \Pr(\shat,\ihat\mid\s,\bmi).
\ee

For a single object, marginalising over latent variables $z$, $\theta$, $s$,
and $i$ in the hierarchical approach is equivalent to convolution of the
error distribution with the model (an example of this is \eqref{eq:FP-conv},
which yields \eqref{eq:FP-ML}). As we show below the marginalisation over $\V$,
$\s$ and $\bmi$ can be performed analytically, while marginalisation over
$\dA$ and $\z$ is straightforward.

The second term on the RHS of \eqref{eq:P-joint-all} is the prior through
which the cosmological model enters. This will be the main focus below. We will
return to the first term in the end to convolve with the error distribution
when we marginalise over $\s$ and $\bmi$.

Now by repeatedly using the chain rule we have
\begin{align}
\Pr(\calC,\V,\r,\s,\bmi,\bm\theta,\dA,\z)
&=\Pr(\r,\s,\bmi \mid \calC,\V,\bm\theta,\dA,\z) \,
	\Pr(\dA \mid \calC,\V,\bm\theta,\z) \,
	\Pr(\calC,\V,\bm\theta,\z) \nonumber\\[3pt]
&=\Pr(\r,\s,\bmi \mid \calC,\bm\theta,\dA) \,
	\Pr(\dA \mid \calC,\V,\z) \,
	\Pr(\V\mid \z,\calC) \,
	\Pr(\z) \, \Pr(\bm\theta) \, \Pr(\calC),
\end{align}
where in the second line we conditioned only on directly related
variables (see Fig.~\ref{fig:FP}).

The fully marginalised posterior to be evaluated is
\be
\begin{split}
\Pr(\calC\mid\bm\zhat,\thhat,\shat,\ihat)
&\propto \Pr(\calC)
\int \dif\V\,\dif\r\,\dif\s\,\dif\bmi\,\dif\dA\,\dif\z\,
	\dif\bm\theta \:\:
\Pr(\zhat\mid\z)\, \Pr(\thhat\mid\bm\theta) \, 
			\Pr(\shat,\ihat\mid\s,\bmi) \\[4pt]
&\qquad\qquad\times\Pr(\r,\s,\bmi \mid \calC,\bm\theta,\dA) \,
	\Pr(\dA \mid \calC,\V,\z) \,
	\Pr(\V\mid \z,\calC) \, \Pr(\z) \, \Pr(\bm\theta).
\end{split}
\ee
This integral can be simplified if we note that
\be
\Pr(\dA \mid \calC,\V,\z) = \delta_D\big(\dA- \dA(\z,\V,\calC)\big)
\ee
and adopt uniform priors $\Pr(\z)=\mrm{const}$ and
$\Pr(\bm\theta)=\mrm{const}$.%
\footnote{Alternatively, the prior may be chosen
based on knowledge of the survey's redshift distribution:
\be
\Pr(z)\,\dif z\propto n(z)\,z^2\,\dif z,
\ee
where $n(z)$ is the redshift distribution. This is auxiliary information unrelated
to the distance indicator itself and evokes the ``orthogonal'' criteria
for constraining distances discussed in \cite{Willick:1994}.
Regardless of what form we choose for $\Pr(\z)$ it is irrelevant for
parameter estimation because of the delta function and the fact that it
depends on survey geometry. More generally, provided the redshift errors are small,
meaning the data are highly informative, the prior should not play a major role.}
Furthermore, we assume for spectroscopic redshift $\hat{z}$ and angular
size $\hat\theta$ that errors are negligible (especially compared with
the $\sim20\%$ distance errors), so that we can make the
following assignments:
\begin{align}
\Pr(\zhat\mid\z) &= \delta_D(\zhat-\z), \\
\Pr(\thhat\mid\bm\theta) &= \delta_D(\thhat-\bm\theta).
\end{align}
However, Gaussian errors on $\hat{z}$ may also be accommodated within this
framework with small modification. In this case to perform the marginalisation
over $\z$ analytically we use that the redshift errors are small and linearise
the angular diameter distance about $z=\hat{z}$; this was done in the hierarchical
model of \cite{March:2011}, finding that neglecting redshift errors do not have a
significant impact on inference.

Absorbing the uniform priors into the proportionality constant and
performing three trivial integrations, we are left with the more
manageable integral
\be\label{eq:post-man}
\Pr(\calC\mid\bm\zhat,\thhat,\shat,\ihat)
\propto \Pr(\calC)
\int \dif\V\,\dif\r\,\dif\s\,\dif\bmi\:\,
	\Pr\big(\r,\s,\bmi \mid \calC,\thhat,\dA(\zhat,\V,\calC)\big) \,
	\Pr\big(\V\mid \zhat,\calC\big) \,
	\Pr(\shat,\ihat\mid\s,\bmi).
\ee
Here we have the integral over the product of three terms.
The third term of the integral is the Gaussian error distribution.
The other two terms we will manipulate into forms that are readily
integrated. In the following sections we show how to analytically perform
the rest of the marginalisations over $\V$, $\r$, $\s$, and $\bmi$.
The basic strategy is to separate out $\r$ from the joint likelihood using
the chain rule. This results in the product of two terms: an $N$-dimensional
Gaussian that depends on $\V$ and a $2N$-dimensional Gaussian that does not.
After marginalising over $\V$ and $\r$ we rearrange the remaining quadratic
forms into a single quadratic form that can be integrated analytically.

\subsection{Developing the terms}
\subsubsection{Fundamental Plane}
First we note the probability of obtaining $r$ given $\theta$ and $d_A$
is non-zero only when $r=\log\theta+\log d_A$; i.e.\
\be
\Pr(\r,\s,\bmi \mid \calC,\thhat,\dA)
=\Pr(\r,\s,\bmi \mid \calC,\thhat,\dA) \,
	\delta_D\big(\r-\bm{\logl_\theta}-\bm{\logl_d}\big),
\ee
where we defined
\begin{subequations}
\begin{align}
&\bm{\logl_\theta} \equiv
\big(\log\hth_1, \log\hth_2, \ldots, \log\hth_N\big)^\T, \label{eq:log_theta}\\
&\bm{\logl_d} \equiv
\big(\log d_{A}(\hat{z}_1), \log d_{A}(\hat{z}_2), \ldots, \log d_{A}(\hat{z}_N)\big)^\T.
\label{eq:log_dA}
\end{align}
\end{subequations}
Marginalisation over $\r$ results in the replacement of $\r$ with
$\bm{\logl_\theta}+\bm{\logl_d}$.

Now, recall that for a single object the FP properties $r,s,i$ are independently
and identically drawn from same underlying population model \eqref{eq:FP-ML}:
\be
r,s,i \sim \mathcal{N}\big((\br,\bs,\bi), \cov^\mrm{FP}\big).
\ee
The individual likelihood $\Pr(r,s,i\mid \calC,\theta,d_A)$ is therefore
a trivariate Gaussian with mean $(\br,\bs,\bi)^\T$ and covariance
$\cov^\mrm{FP}=\mat{O}\mat{D}\mat{O}^\T$.
While the joint likelihood \eqref{eq:lnL-FP} can be written as the product
of $N$ trivariate Gaussians, to facilitate integration we will instead
form a $3N$-dimensional multivariate Gaussian for which the first $N$ rows
and columns correspond to $\r$, the next $N$ correspond to $\s$, and the
last $N$ correspond to $\bmi$. In particular
\be
\r,\s,\bmi \sim \mathcal{N}\big((\rbar,\sbar,\ibar),\mathring{\cov}\big),
\ee
where the joint covariance is partitioned in block form as
\be\label{eq:Co}
\mathring\cov
\equiv\begin{pmatrix}
\Crr & \CryT \\
\Cry & \Cyy
\end{pmatrix}.
\ee
Here $\Crr$ is a $N\times N$ matrix, $\Cry$ is a $2N\times N$
matrix, and $\Cyy$ is a $2N\times 2N$ matrix.
In this way, when conditioning on $\s$ and $\bmi$, we may use the formulae
of Appendix \ref{app:cond-gauss}. Since we will be marginalising over $\V$
we require the conditional form
\be\label{eq:rsi}
\Pr(\r,\s,\bmi\mid \calC,\bm\theta,\dA)
=\Pr(\r\mid \s,\bmi,\calC,\bm\theta,\dA) \,
	\Pr(\s,\bmi\mid\calC),
\ee
where we have dropped the conditioning on $\bm\theta$ and $\dA$ in the
second term.
The first term is
\be
\Pr(\r \mid \s,\bmi,\calC,\bm\theta,\dA)
=\calN(\r\, ;\, \rbar',\Crr')
\ee
with (see Appendix \ref{app:cond-gauss})
\begin{subequations}
\label{eq:shifted-r-moments}
\begin{align}
&\rbar' 
	= \rbar - \CryT \, \Cyyi \, (\ybar-\y), \label{eq:rbar-prime} \\
&\Crr' = \Crr - \CryT \, \Cyyi \, \Cry,
\end{align}
\end{subequations}
where $\y\equiv(\s,\bmi)^\T$ and $\ybar\equiv(\bm\bs,\bm\bi)^\T$ are
$2N$-dimensional vectors.
The second term of \eqref{eq:rsi} is given by a higher-dimensional analog of
\eqref{eq:FP-ML}, marginalised
over $\r$ (i.e.\ striking out the first $N$ rows and columns):
\be\label{eq:p-si}
\Pr(\s,\bmi\mid\calC) = \calN(\y\, ;\, \ybar,\Cyy).
\ee
Altogether we have
\be\label{eq:rsi-prod}
\Pr(\r,\s,\bmi \mid \calC,\thhat,\dA)
=\calN(\r\, ;\, \rbar',\Crr')\,
\calN(\y\, ;\, \ybar,\Cyy) \,
\delta_D\big(\r-\bm{\logl_\theta}-\bm{\logl_d}\big).
\ee
Finally we have for the error distribution
\be\label{eq:p-si-hatted}
\Pr(\shat,\ihat\mid\s,\bmi) = \calN(\yhat\, ;\, \y, \Eyy),
\ee
where $\Eyy$ is constructed in a similar way to $\Cyy$ of \eqref{eq:Co}.

\subsubsection{Distance-redshift relation}
At the low redshifts typical of a peculiar velocity catalogue the
angular diameter distance is given by (see \cite{Hui:2005nm}; for a
direct calculation see \cite{Kaiser:2014jca})
\be\label{eq:dA-perturbed}
d_A(z)=\bar{d}_A(z)(1-\kappa),\qquad
\kappa(z)=\left[1-\frac{d_H(z)}{\bar{d}_A(z)}\right]\frac{v}{c},
\ee
where $d_H(z)=c/H(z)$ and all terms are evaluated at the total
redshift $z=\hat{z}$.%
\footnote{The difference $r -\log(\theta \bar{d}_A)$ is not the
same as $\Delta r$ in \cite{Springob:2014qja}; $\bar{d}_A$ here is
evaluated at $z$ \emph{not} $\bar{z}$.} It is in this regime that the
dominant contribution to the convergence $\kappa$ is from peculiar velocity.
Since $v/c\sim10^{-3}$ we have $\kappa\lesssim 0.1$ for $z\gtrsim 0.01$, and
we therefore approximate $\ln(1-x)\simeq -x$, as in \cite{Adams:2017val},
and write
\be
\log d_A(z) \simeq \log\bar{d}_A(z) - \frac{\kappa(z)}{\ln10},
\ee
or in vector form
\be\label{eq:l-dA}
\bm{\logl_d} = \bm{\logl_{\bar{d}}} + \A\V,
\ee
where we defined
$\bm{\logl_{\bar{d}}}
=\big(\log\bar{d}_A(\hat{z}_1),\log\bar{d}_A(\hat{z}_2),\ldots,\log\bar{d}_A(\hat{z}_N)\big)^\T$
and the $N\times N$ symmetric matrix
\be
\A=\mrm{diag}(A_1,A_2,\ldots,A_N),\qquad
A_m = 
    \frac{1}{c\ln 10}
    \left[\frac{d_H(\hat{z}_m)}{\bar{d}_A(\hat{z}_m)}-1\right],
\ee
and we note that $A_m\sim 1/z_m$ at low redshifts, and is why the fluctuations
to the distance can be large.

\subsubsection{Large-scale structure\label{sec:LSS}}
As well as the usual distance-redshift relation, cosmology
also enters through correlations in the source velocities.
Because neighbouring sources will move with similar velocity, we expect
correlations between source pairs meaning the PDF of $\V$ will not
be separable. In linear theory, the joint peculiar velocity distribution
is described by a multivariate Gaussian
\be\label{eq:pec-vel-like}
\Pr(\V \mid \z,\calC)
=\calN(\V\, ;\, \bm0,\R)
=\frac{1}{\det(2\pi\R)^{1/2}}
	\exp\left(-\frac{1}{2}\V^\T \R^{-1} \V\right).
\ee
Here the prior of $\V$ is taken to be the likelihood in the standard
analysis taken over the catalogue data
(e.g.\ \cite{Jaffe:1994gx,Ma:2010ps,Macaulay:2011av,Johnson:2014kaa}).
Depending on the model under consideration other choices of prior are
possible. However, for the current discussion we will focus on a spatially
flat \LCDM\ cosmology. The covariance between the $m^\mrm{th}$ and $n^\mrm{th}$
source is given by
\be\label{eq:R}
R_{mn}
=\big\langle v(\bm{x}_m)\, v(\bm{x}_n)\big\rangle
=\xi_v(\bm{x}_m,\bm{x}_n),
\ee
where $\xi_v$ is the two-point correlation function of the LOS
velocities. The $m^\mrm{th}$ source has a redshift-space position of
$(z_m,\alpha_m,\delta_m)$, or in real-space
$\bm{x}_m=D_m\,\n_m$, with $D_m\equiv D(z_m)$ the comoving distance and
$\n_m=\n(\alpha_m,\delta_m)$ the direction of observation. Notice that we
set the galaxy distance at their \emph{observed} redshift $z$, and not the
background redshift $\bar{z}$, corresponding to the real-space position.
Just like redshift-space density fluctuations, peculiar velocities are
also affected by redshift-space distortions; however, in linear theory
the real-space and redshift-space velocity correlation functions are equivalent
\citep{Koda:2013eya,Okumura:2014}. We reiterate that in this work we only
demonstrate a template analysis from which we can develop more sophisticated
models.

On subhorizon scales, typical of peculiar velocity surveys, we can use the
linearised continuity equation ${\delta'+\nabla\cdot\v=0}$ so that the
correlation function can be expressed in terms of the matter power spectrum.
For production work, the correlation function is better formulated in terms
of the velocity divergence $\nabla\cdot\v$
because it does not manifestly depend on the linearised continuity equation
and require a galaxy bias model; non-linear corrections are also more easily
implemented \citep{Johnson:2014kaa}. The correlation function reads
\be
\xi_v(\bm{x}_m,\bm{x}_n)
=\frac{1}{2\pi^2}\,f^2_0\,\HO^2\int{\dif k}\, W_{mn}(k) P(k),
\ee
where, at the low redshifts typical of peculiar velocity surveys, we have made
the usual assumption about equal-time correlations; $f_0\equiv f(a_0)$ is
the present-day growth rate, with $f_0\approx\Omega_m^\gamma$, and $\gamma\approx0.55$
for \LCDM. We have also the matter power spectrum  $P(k)$, and the window function
\begin{align}
W_{mn}(k)
&\equiv\int \frac{\dif^2 {\hat{\k}}}{4\pi}\, e^{-ik\hat{\k}\cdot(\bm{x}_m-\bm{x}_n)}
		(\hat{\k}\cdot\n_m)(\hat{\k}\cdot\n_n) \nonumber\\
&=\frac{1}{3}\big[j_0(kr_{mn})-2j_2(kr_{mn})\big]\cos\Upsilon_{mn}
    +\frac{D_m D_n}{r_{mn}^2}j_2(kr_{mn})\sin^2\Upsilon_{mn}.
\end{align}
Here $r^2_{mn}=|\bm{x}_m-\bm{x}_n|^2
=D^2_m+D^2_n-2D_mD_n\cos\Upsilon_{mn}$ by the cosine rule, and
$\Upsilon_{mn}=\arccos(\n_m\cdot\n_n)$ is the angular separation between
the $m^\mrm{th}$ and $n^\mrm{th}$ object; $j_0$ and $j_2$
are the zeroth and second order spherical Bessel functions, respectively.
The second line is expressed in terms of observer-centric quantities [see
\cite{Ma:2010ps} for a derivation]; it is equivalent to the more common
decomposition in terms of parallel and perpendicular kernels
(see, e.g.\ \cite{Peebles:1993}).

It is also necessary to include in $\R$ a velocity dispersion $\sigma^2_*$ so
that $R_{mn}\to  R_{mn} + \sigma^2_*\, \delta_{mn}$
in which $\sigma_*$ is treated as a free parameter included in $\calC$.
This parameter captures the one-dimensional incoherent Gaussian random motion
of galaxies on non-linear scales.
We remark that there is a slight difference with the usual approach, which is
that here $v_m$ are the latent radial velocities so have no catalogue error.%
\footnote{In the standard approach in which peculiar velocities are given with
some uncertainty we would have
$R_{mn}\rightarrow (R_{mn}+\sigma^2_m\delta_{mn})+\sigma^2_*\delta_{mn}$,
where $\sigma_m$ is the uncertainty on the $m^\mrm{th}$ source's peculiar velocity.}

\subsection{Marginalisation}
After inserting \eqref{eq:rsi-prod}, \eqref{eq:l-dA}, and
\eqref{eq:pec-vel-like} into \eqref{eq:post-man} and rearranging slightly,
the posterior reads
\be
\begin{split}
\Pr(\calC\mid\bm\zhat,\thhat,\shat,\ihat)
\propto \Pr(\calC)
&\int\dif\y\: \calN(\yhat\, ;\, \y,\Eyy) \, \calN(\y\, ;\, \ybar,\Cyy) \\
&\times\int\dif\V\int\dif\r\:
		\calN(\r\, ;\, \rbar',\Crr')\,
		\delta_D\big(\r-\bm{\logl_\theta}-\bm{\logl_{\bar{d}}}-\A\V\big) \,
		\calN(\V\, ;\, \bm0,\R).
\end{split}
\ee
Integrating out $\r$ is trivial because of the delta function, and gives
\be
\begin{split}
\Pr(\calC\mid\bm\zhat,\thhat,\shat,\ihat)
\propto \Pr(\calC)
&\int\dif\y\: \calN(\yhat\,;\, \y, \Eyy) \, \calN(\y\, ;\, \ybar,\Cyy) \\
&\times\int\dif\V\:
	\calN\big(\rbar'\, ;\, \r(\bm{\logl_\theta},\bm{\logl_{\bar{d}}},\A\V,\calC),\Crr'\big)\,
		\calN(\V \, ;\,  \bm0,\R),
	\label{eq:post-double-int}
\end{split}
\ee
where
$\r(\bm{\logl_\theta},\bm{\logl_{\bar{d}}},\A\V,\calC)
=\bm{\logl_\theta} + \bm{\logl_{\bar{d}}} + \A\V$.
The above expression can be thought of as a double convolution:
the first convolves the distance-dependent part of the FP (for a given
distance-redshift relation) with the peculiar velocity distribution due
to correlations from large-scale structure and cosmic variance; the
second convolution is with the distance-independent part of the FP.
Note that the two integrals cannot be separated because
$\rbar'$ depends on $\y$ by \eqref{eq:rbar-prime}.

With a change of variables $\V\to\U=\A\V$ the inner $\V$ integral of
\eqref{eq:post-double-int} is readily performed using \eqref{eq:gauss-conv}
to give
\begin{align}
\Pr(\calC\mid\bm\zhat,\thhat,\shat,\ihat)
\propto \Pr(\calC)
&\int\dif\y\: \calN(\yhat\, ;\, \y,\Eyy) \, \calN(\y\, ;\, \ybar,\Cyy) \,
    \calN(\bm\Delta_r\, ;\, \bm0,\Sigrr),
\end{align}
where we defined
\begin{subequations}
\begin{align}
&\bm\Delta_r
\equiv \rbar'- \bm{\logl_\theta} - \bm{\logl_{\bar{d}}}
= \big[\bm{\br} - \CryT \, \Cyyi\,(\ybar-\y)\big]
				- \big(\bm{\logl_\theta} + \bm{\logl_{\bar{d}}}\big),  
		\label{eq:Delta-r}\\[4pt]
&\Sigrr
\equiv \A\R\A + \Crr'
=\A\R\A	+ \Crr	- \CryT \, \Cyyi \, \Cry \,.
	\label{eq:Sig-rr}
\end{align}
\end{subequations}
This leaves us with one final integral, which can be done by bringing the
integrand into Gaussian canonical form then using \eqref{eq:gauss-int2}. The
details of this calculation are given in Appendix~\ref{app:details}; here we
state only the final result:
\begin{align}
\Pr(\calC\mid\zhat,\thhat,\shat,\ihat)
\propto
\frac{1}{\det\mbf{\Sigma}^{1/2}}
\bigg[\prod_{m=1}^N\frac{1}{\det(\cov+\E_m)^{1/2}}\bigg]
\exp\bigg[
    -\frac{1}{2}\bm\Delta^\T\,\mbf{\Sigma}^{-1}\,\bm\Delta
	-\frac{1}{2}\sum_{m=1}^N\Delta\y_m^\T\,(\cov+\E_m)^{-1}\,\Delta\y_m
	\bigg]\, \Pr(\calC),
\label{eq:post}
\end{align}
where
\begin{subequations}
\begin{align}
\bm\Delta
&=\big[\rbar - \CryT\,(\Cyy+\Eyy)^{-1}\,(\ybar-\yhat)\big]
            - (\bm{\logl_\theta} + \bm{\logl_{\bar{d}}}), \label{eq:Delta}\\[4pt]
\Sig &=\A\R\A+\Crr-\CryT\,(\Cyy+\Eyy)^{-1}\,\Cry.
\end{align}
\end{subequations}
and $\Delta\y_m=(\hs_m-\bs,\hi_m-\bi)^\T$, with $\cov$ and $\E_m$ being the
corresponding $2\times2$ submatrices of $\cov^\mrm{FP}$ and $\E^\mrm{FP}_m$,
respectively. It can be seen that the joint posterior density is composed of two
Gaussian densities (that we have written into a single exponential):
The first is cosmological in nature, accounting for the distance-redshift
relation and cosmic variance; the second is purely related to the physical
characteristics of galaxies. Notice that the presence of $\R$ correlates
all galaxies; this is in contrast to the conventional analysis, which
considers only correlations from the FP. Here $\Sig$ is a
dense matrix because of the presence of $\A\R\A$; the first quadratic
form in the exponential of \eqref{eq:post} cannot be reduced down to
the product of smaller terms.
By comparison the physical properties of each galaxy (velocity dispersion
and surface brightness) being independent of one another allows us
to write the $2N\times 2N$ quadratic form of $\yhat$ as the sum of
$N$ $2\times2$ quadratic forms, c.f.\ \eqref{eq:FP-conv}.

Except for $\Eyy$, note that all matrices depend on parameters
so that the determinants must be included in any parameter scans.
We further emphasize $\bm{\logl_{\bar{d}}}$ also depends on parameters
through the angular diameter distance.

Aside from the factors of $2\pi$ we have omitted, the proportionality
also accounts for prior on $\zbar$, which is unimportant for parameter
estimation when the uncertainties are assumed to be negligible.

\subsection{Recovering the Fundamental Plane likelihood}
As a consistency check, we verify that the standard FP likelihood
\eqref{eq:FP-ML} can be recovered if we fix the cosmological parameters
and take $\R\to0$ (no correlations from large-scale structure).\footnote{This is
equivalent to having assigned a Dirac delta function prior for the peculiar
velocities centered at zero, because
\be
\Pr(\V\mid\z,\vartheta)=
\lim_{\R\to0} \:
\frac{1}{{\det(2\pi\R)}^{1/2}}
	\exp\left(-\frac{1}{2}\V^\T \R^{-1} \V\right)
=\delta_D(\V).
\ee
}
Now, as the mapping from distance to physical size is fully determined
by the (known) cosmological parameters, we can swap the observables
$\bm{\logl_\theta}$ and $\bm{\logl_{\bar{d}}}$ with the conventional
size observable defined as $\bm\rhat\equiv\bm{\logl_\theta}+\bm{\logl_{\bar{d}}}$.
We thus have $\bm\Delta=\rbar'-\bm\rhat$, with
$\rbar'\equiv \rbar-\CryT(\Cyy+\Eyy)^{-1}(\ybar-\yhat)$ the shifted mean.
This recovers \eqref{eq:FP-ML} in conditional form
\begin{align}
\Pr(\calC\mid\bm\rhat,\shat,\ihat)
&\propto
\frac{1}{\det\Sig^{1/2}}
	\exp\left[
	    -\frac{1}{2}(\rbar'-\bm\rhat)^\T\,\Sig^{-1}\,(\rbar'-\bm\rhat)
	    \right]
\prod_{m=1}^N\frac{1}{\det(\mbf{C}+\mbf{E}_m)^{1/2}}
\exp\left[
	-\frac{1}{2}\Delta\y_m^\T\,(\mbf{C}+\mbf{E}_m)^{-1}\,\Delta\y_m
	\right].
\end{align}
Without correlations induced by $\R$ we have that
$\Sig=\Crr-\CryT\,(\Cyy+\Eyy)^{-1}\,\Cry$ is a diagonal matrix,
allowing the first two terms to be factorised into a product
of $N$ univariate Gaussians. The resulting expression can thus be
manipulated into the form of the product of $N$ trivariate Gaussians.

\begin{figure}
   \centering
   \includegraphics[scale=0.42]{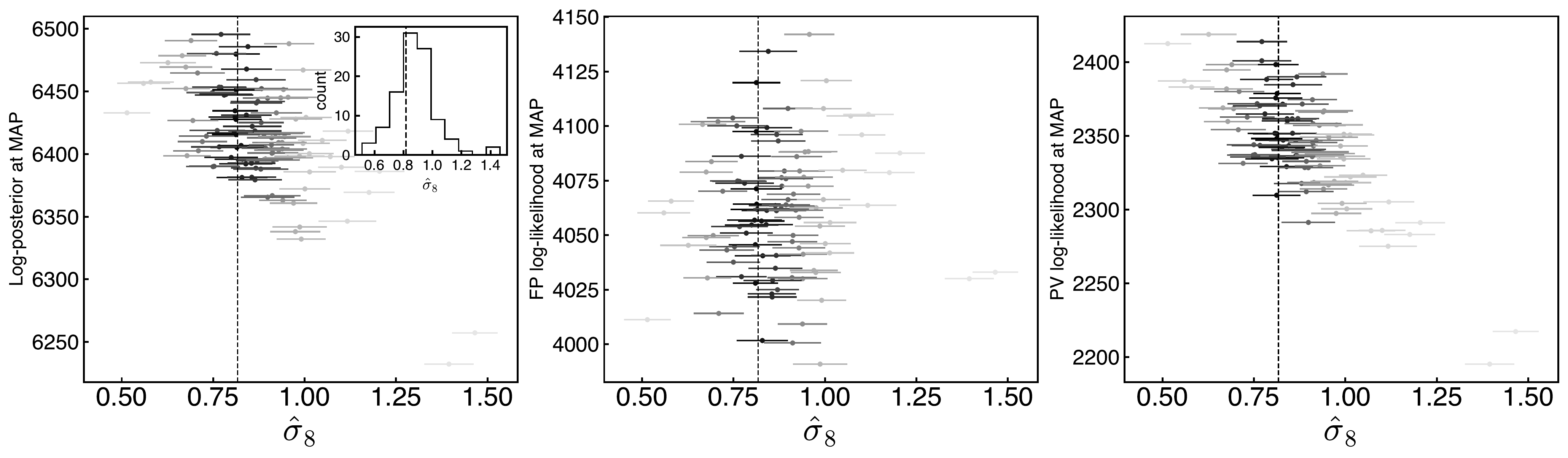}
   \caption{Maximum a posterior estimates of $\sigma_8$ from 100 FP mock data realisations
            of $N=1000$ galaxies and plotted against the value of the logarithm of the
            likelihood or posterior. As a crude estimate of the uncertainty we also show
            the associated Hessian errors of $\sigma_8$. The gray-scale indicates
            the RMS difference from the true value $\sigma_8=0.817$ (dashed line),
            with the darkest
            having the smallest RMS and lightest the largest. Note the difference
            between the sum of
            the FP and peculiar velocity log-likelihoods and the log-posterior is because of
            the log-uniform priors we assign to the scale parameters $\sigma_1$,
            $\sigma_2$, $\sigma_3$, and $\sigma_8$; all other parameters are assigned
            uniform priors.}
   \label{fig:sims}
\end{figure}

\begin{figure}
  \centering
  \includegraphics[scale=0.35]{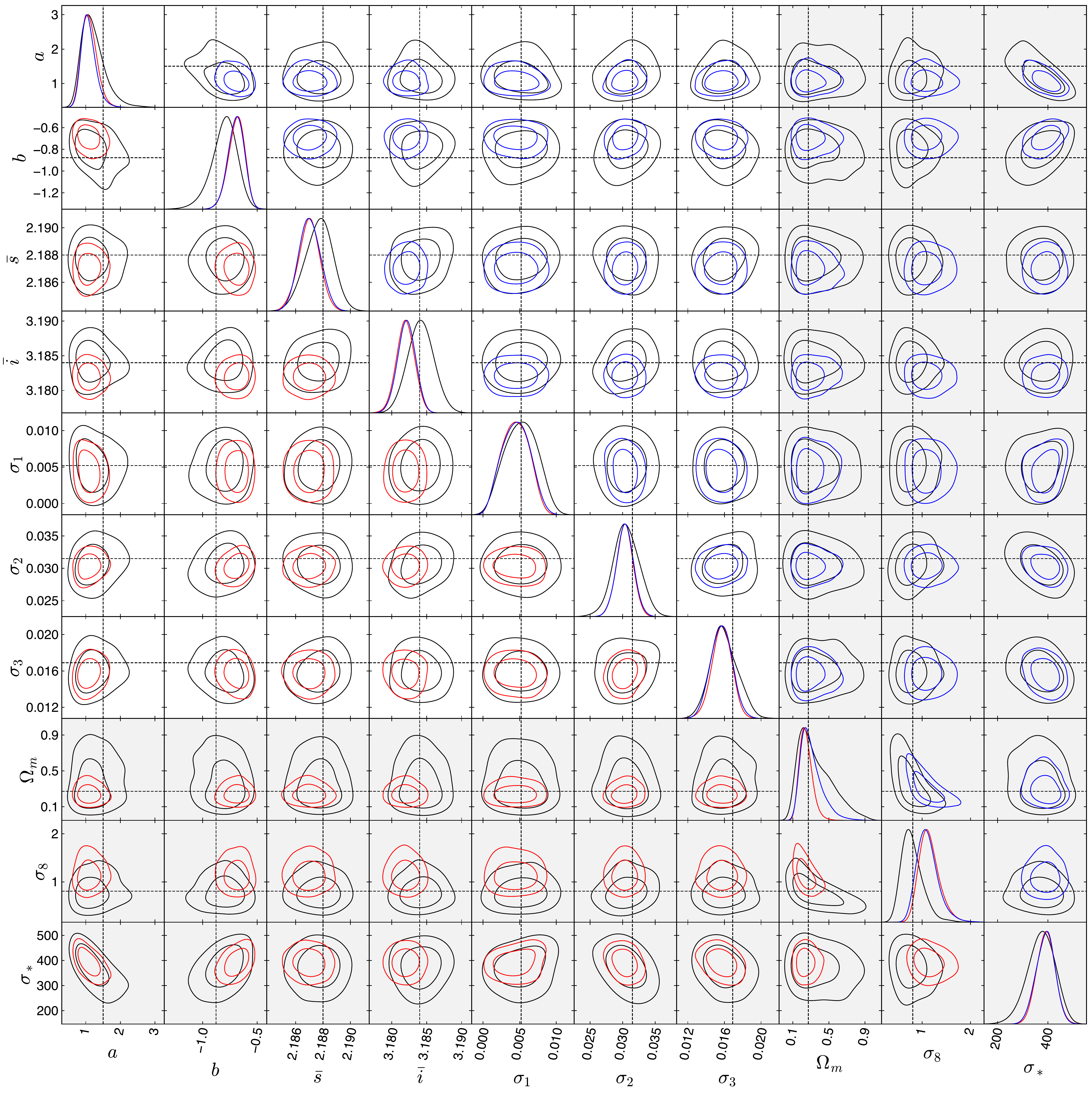}
  \caption{The marginalised posteriors when (\emph{i}) $\br$ is free to vary (black curves),
            (\emph{ii}) the calibration is fixed to a value $10\%$ lower than the true value
            $\br=0.191$ (blue curves), and (\emph{iii}) the calibration is fixed to a value $10\%$
            higher than the true value (red curves).
            Note in each case all other calibration parameters are fitted for.
            The true values are indicated by dashed black lines.
            Particular attention should be paid to joint contours between cosmological
            and FP parameters (highlighted in grey), where calibration parameters fixed
            to wrong values can potentially lead to systematic shifts in the parameters of
            interest. These constraints
            should not be viewed as indicative of the performance on realistic data
            sets as here we fit to a small sample of galaxies ($N=1000$) and assume
            $1\%$ statistical errors on $\hs$ and $\hi$. Here $\sigma_*$ is in units
            $\mrm{km/s}$ and is not known a priori.
            }
  \label{fig:triplot-all}
\end{figure}

\subsection{Fundamental Plane calibration uncertainty\label{sec:uncertain}}
Another advantage of our unified treatment is that the uncertainty in
FP calibration parameters can be straightforwardly propagated downstream to the
cosmological parameters. Although
we have left the nuisance parameters (i.e.\ FP parameters) unmarginalised,
the centroid parameters $\br$, $\bs$, and $\bi$ can in fact be analytically
marginalised over if we assume Gaussian priors. For example, carrying out the
marginalisation assuming a Gaussian prior on $\br$ with mean $\mu_{\br}$
and variance $\sigma^2_{\br}$, the final result is a modified form
of \eqref{eq:post} with $\bm\rbar=\br\mat{1}\to\mu_{\br}\mat{1}$, and a
monopole contribution to the covariance,
$\Sigrr\to\Sigrr+\sigma_{\br}^2\,\mat{J}_N$, where $\mat{J}_N$ is the
$N\times N$ matrix of ones \citep{Bridle:2001zv}.
This shows that uncertainty in global parameters like
$\br$ induces an ambient error and covariance for all objects.

We note that we can also analytically marginalise over $\bs$ and $\bi$
in a similar way, but that the parameters $a$ and $b$ enter into the
covariance matrices so will have to be marginalised over by other means.

\subsection{Maximum a posteriori estimator for peculiar velocities}
The posterior \eqref{eq:post} derived is marginalised over all
peculiar velocities. However, if we leave $\V$ unmarginalised then we would have an expression for
$\Pr(\vartheta,\V\mid\mathscr{D})$, with $\mathscr{D}=\{\zhat,\thhat,\shat,\ihat\}$,
and the posterior for $\V$ is
\be\label{eq:V-post}
\Pr(\V\mid\mathscr{D}) = \int\dif\vartheta\, \Pr(\vartheta,\V\mid\mathscr{D}).
\ee
Building probabilistic peculiar velocity catalogues from \eqref{eq:V-post} would be desirable from
a Bayesian perspective as any uncertainty in the calibration and cosmology
is accounted for \citep{Brewer:2012gt,Portillo:2017}.
The problem, however, is that \eqref{eq:V-post} is a high-dimensional posterior
and standard MCMC methods are unfeasible for realistic data sets containing
$N\gtrsim 10^4$ galaxies and clusters, though Gibbs sampling can be
effective provided the conditional distributions of the target posterior
have a simple form.\footnote{For example, \cite{Alsing:2016} has developed a hierarchical
model for cosmic shear maps based on $\sim250,000$ parameters, along with an efficient
Gibbs sampling scheme.} An alternative is to construct point
estimators for $\V$. Since the prior on $\V$ is a multivariate Gaussian
the joint posterior containing $\V$ is the product of two Gaussians.
It is straightforward to solve
${\partial\Pr(\calC,\V\mid\mathscr{D})/\partial\V=0}$ for $\V$
to obtain the maximum a posteriori (MAP) estimate
\be
\widehat{\V}_\mrm{MAP}
=\R\A\Sig^{-1}\bm\Delta.
\ee
Manipulating this slightly, the estimator can be seen to be equivalent to
the Wiener filtering of the residuals $\A^{-1}\bm\Delta$:
\be\label{eq:VMAP}
\widehat{\V}_\mrm{MAP}
=\big[\R(\R+\A^{-1}\Sig_0\A^{-1})^{-1}\big] \A^{-1}\bm\Delta.
\ee
where $\Sig_0\equiv\Crr-\CryT\,(\Cyy+\Eyy)^{-1}\,\Cry$, which differs
from $\Sig$ by $\A\R\A$, and the term in square brackets is the linear
optimal filter with signal $\R$ and noise $\A^{-1}\Sig_0\A^{-1}$. The
associated covariance of the MAP estimate is given by the inverse of
the Hessian of the log-posterior:
\be\label{eq:CMAP}
\widehat{\cov}_\mrm{MAP}
= (\R^{-1} + \A\Sig_0^{-1}\A)^{-1}.
\ee

As discussed in Section \ref{sec:uncertain}, uncertainty in the
calibration parameters can be built into the estimate. If we are
uncertain about the zero-point (i.e.\ $\br$) by $\sigma_\mrm{ZP}$,
then we have the slight modification
\be
\widehat{\cov}_\mrm{MAP}
= \big[\R^{-1} + \A(\sigma^2_\mrm{ZP}\,\mat{J}_N + \Sig_0)^{-1}\A  \big]^{-1},
\ee
where $\mat{J}_N$ is the $N\times N$ matrix of ones.

\section{Numerical experiments}\label{sec:expts}
As a proof of concept, we analyse mock data with the
aim of recovering the true parameters given some measurement error. Recall we
require the following data:
\be
\mathscr{D}
=\Big\{\big(\hat{z}_m,\,\hth_m,\,\hs_m,\,\hi_m,\,\hat{\alpha}_m,\,\hat{\delta}_m\big)\Big\}_{m=1}^N.
\ee
The positions and peculiar velocities of the galaxies are fixed according to a halo
catalogue we obtain from the Big MultiDark Planck $N$-body simulation (BigMDPL) 
\citep{Klypin:2016}.
BigMDPL is a dark matter only simulation performed using the {\sc l-gadget-2} code
\citep{Springel:2005}. The simulation box has a side length of $2.5h^{-1}\,\mrm{Gpc}$,
a mass resolution of $2.4\times 10^{10}h^{-1}\, M_\odot$, and consists of
$3840^3$ particles, which are evolved forward from an initial redshift of $z=100$.
The simulation assumes a spatially flat \LCDM\ cosmology with $h=0.678$,
$\Omega_m=0.307$, $\Omega_\Lambda=0.693$, $\Omega_b=0.048$, $n_s=0.96$,
and $\sigma_8=0.829$. 
We use the halo catalogue derived from the $z=0$ snapshot, which was
constructed using the {\sc rockstar} halo finder \citep{Behroozi:2013}.
Halos are selected in the mass range of ${10^{11.5}-10^{12}}h^{-1}\, M_\odot$.
Simulation data is obtained from the CosmoSim online
database.\footnote{\url{https://www.cosmosim.org/}}

The FP data are generated using $a=1.502$, $b=-0.877$, $\br=0.191$, $\bs=2.188$,
$\bi=3.184$, $\sigma_1=0.0052$, $\sigma_2=0.0315$, and $\sigma_3=0.0169$.
In addition we also include the free parameter
$\sigma_*$ to capture the small scale random motions not described by linear theory.
In all fits we fix $n_s$, $h$, and $\Omega_b$ to their true values.

We draw $N=1000$ triples $(r_m,s_m,i_m)$ from the intrinsic population
given by the trivariate Gaussian with mean $\xbar$ and covariance $\cov$, i.e.\ 
$\x_m\sim\mathcal{N}(\xbar,\cov)$. The selection criteria of our halo catalogue
is chosen so that it follows a 6dFGS-like distribution of galaxies
\citep{Campbell:2014uia}. Thus sky positions will be located in one hemisphere
relative to an observer placed in the centre of the simulation box. The redshift
distribution is chosen such that number density is approximately constant, so the differential
number density is $(\dif n/\dif z)\,\dif z\propto z^2\,\dif z$ in the range $z$ to $z+\dif z$,
and bounded by $z_\mrm{min}=0.006$ and $z_\mrm{max}=0.05$. The observed redshift $z_m$
is computed from $z_m=(1+\bar{z}_m)(1+v_m/c)-1$, with $\bar{z}_m$ obtained from the
comoving position and $v_m$ taken directly from the halo catalogue.
The perturbed angular diameter distance $d_A$ as a function of the observed
redshift $z_m$ and peculiar velocity $v_m$ is computed using \eqref{eq:dA-perturbed},
giving an angular size of $\theta_m=10^{r_m}/d_A(z_m)$.

Our choice of analysing a relatively small number of galaxies is made for reasons
of speed. Realistic FP samples have $N\gtrsim10^4$, requiring the inversion of
matrices of size $N\times N$, which is a significant computational cost.
For analyses of real data, a practical solution is to use a gridding method
\citep{Abate:2008zy,Johnson:2014kaa} as a form of dimensional reduction;
here, as a concession to the small sample size, we will take slightly more
optimistic measurement errors instead.
The errors $\epsilon_{s,m}$ and $\epsilon_{i,m}$ will thus be drawn from
a Gaussian with a standard deviation equal to
$1\%$ of the values of $s_m$ and $i_m$ so that we take as observed quantities
$\hs_m=s_m+\epsilon_{s,m}$ and $\hi_m=i_m+\epsilon_{i,m}$. (For comparison,
6dF galaxy survey errors on $\hi$ in $J$ band and $\hs$ are around the $2-3\%$
level.)

Since $a$, $b$, and $c$ have non-trivial correlations, we performed the sampling
in the space of FP centroids
$\xbar=(\br,\bs,\bi)^\T$
and $3\times3$ covariance matrices $\cov^\mrm{FP}$ parametrized by $\sigma_r$,
$\sigma_s$, $\sigma_i$, and the correlation coefficients $\rho_{rs}$,
$\rho_{ri}$, $\rho_{si}$. These parameters are generally less correlated
and better sampling behaviour. Note there is one additional free parameter than
in the standard FP fit and this corresponds to the freedom to
rotate the FP on to itself, without changing $a$ and $b$. While this does not change
the FP relation, allowing for this rotation does change the quality of fit;
here we do not fix this degree of freedom.
The FP orientation parameters $a$ and $b$ are derived from the MCMC chains as a
post-processing step. These are obtained by computing the eigenvector
$\bm{e}_1=(A,B,C)^\T$ of $\cov^\mrm{FP}$ with the lowest eigenvalue.
This corresponds to the direction with lowest variances and defines the FP;
FP relation parameters are then given by $a=-B/A$ and $b=-C/A$ (see Appendix
\ref{app:geometry}). Moreover, the parameters $\sigma_1$, $\sigma_2$, and $\sigma_3$
can be computed by taking the square root of the eigenvalues of $\cov^\mrm{FP}$.

We assign uninformative priors that are flat in
$\ln\sigma_1$, $\ln\sigma_2$, $\ln\sigma_3$, $\rho_{rs}$, $\rho_{ri}$, $\rho_{si}$,
$\br$, $\bs$, $\bi$, $\Omega_m$, $\ln\sigma_8$, and $\ln\sigma_*$; the correlation
coefficients are bounded by $-1$ and $1$, while $\Omega_m$ is bounded by
0 and 1. The posterior distribution is sampled using an affine-invariant
ensemble MCMC scheme \citep{ForemanMackey:2012ig}.

In Fig.~\ref{fig:sims} we show the robustness of $\hat\sigma_8$
for 100 different data realisations. Note that peculiar velocities are
generated independent of the FP (latent) data $\r,\s,\bmi$. The fact that the
recovered $\hat\sigma_8$ departs from the true value is not because of
the FP data but the particular peculiar velocity sample drawn; for a highly
dispersive sample $\hat\sigma_8$ is larger than the true value (lower
values of the peculiar velocity likelihood), while it is smaller for a
sample drawn near the centroid (higher values of the peculiar velocity
likelihood.

In Fig.~\ref{fig:triplot-all} we show the estimated parameters from
one realisation of the ersatz data set with $N=1000$ galaxies and $1\%$ errors on
the measured $\hs$ and $\hi$.
Though these errors are optimistic -- realistically we can expect $\gtrsim 2\%$
from current catalogues -- we are analysing a much smaller data set.
Also shown is the impact when the zero-point,
$\br$, or equivalently $c$, is fixed. Note in the joint fit ($\br$ free)
we find $0.194^{+0.006}_{-0.010}$ ($68\%$ C.L), very much consistent with the
true value. As expected, the overconfidence can be seen to
produce tighter constraints, but can also cause a systematic shift in other parameters
when $\br$ is fixed away from the mode (i.e.\ $\br$ is biased). This is most
apparent for $\sigma_8$ when $\br$ is biased $\pm10\%$ from the true value.
In the case when $\br$ is a free parameter the constraint on $\Omega_m$ is
considerably degraded; furthermore, the constraint on $\Omega_m$ when $\br$ is
biased high is noticeably stronger than when it is biased low.
To some extent the zero-point marginalisation procedure of
\cite{Johnson:2014kaa} performed post-calibration offsets some of this bias,
but assumes the marginal posterior of $\br$ is exactly Gaussian. The above
asymmetric constraints and biases suggests this can only be a first approximation.
The difference with our consistent marginalisation will be further investigated
in future work.

\section{Selection effects}\label{sec:selection}
The likelihood \eqref{eq:post} derived above applies in an idealised
analysis in which all objects along the LOS may be observed. In practice,
survey instruments have limited sensitivity and only the brightest
objects are seen. These selection effects must be accounted to
ensure unbiased inference. In this section we make some brief remarks
about how these can be included, deferring a more detailed study to future work.

Let the data be denoted $\mathscr{D}=\{\zhat,\thhat,\shat,\ihat\}$ (and possibly
the experimental covariance matrix), and $\calS$ be the proposition
that we have observed some data thus passing the selection criteria.
The probability that a given data set is observed depends on the
selection criteria.
The likelihood of observing $\mathscr{D}$ given $\vartheta$ \emph{and} that we
have observed some data is \citep{Loredo:2004,Mandel:2018mve}
\be\label{eq:p-select}
\Pr(\mathscr{D}\mid\calS,\vartheta)
=\frac{\Pr(\calS\mid\mathscr{D},\vartheta)\,\Pr(\mathscr{D}\mid\vartheta)}
	{\Pr(\calS\mid\vartheta)}
=\frac{\Pr(\calS\mid\mathscr{D},\vartheta)\,\Pr(\vartheta\mid\mathscr{D})}
	{\int\dif\mathscr{D}'\,\Pr(\calS\mid\mathscr{D}',\vartheta)\,
	\Pr(\vartheta\mid\mathscr{D}')},
\ee
where $\Pr(\calS|\mathscr{D},\vartheta)$ is the selection function, and we
used Bayes' theorem in the second equality so
$\Pr(\vartheta|\mathscr{D})
	\propto \Pr(\mathscr{D}|\vartheta)\Pr(\vartheta)$.
If we assume that all objects that exceed some threshold are successfully
observed then $\Pr(\calS|\mathscr{D},\vartheta)=1$. In the ideal scenario of
no selection effects (see Section \ref{sec:bhm}) clearly all possible
data sets are observable. In the case of a cutoff, equivalent to
replacing with a truncated distribution, as in \eqref{eq:FP-ML}, we have
$\Pr(\calS|\mathscr{D},\vartheta)=1$ if it exceeds some threshold, and
$\Pr(\calS|\mathscr{D},\vartheta)=0$ otherwise.
Thus $\Pr(\calS|\vartheta)$ is the fraction of all possible data sets that are
observable given parameters $\vartheta$.
From \eqref{eq:p-select} we can see that the form of the posterior distribution
subject to selection effects, $\Pr(\vartheta|\mathscr{D},\calS)$, is simply
that of $\Pr(\vartheta|\mathscr{D})$ given by \eqref{eq:post} but with a
different normalisation.

As regards the FP there are two selection criteria to consider:
(\emph{i}) The spectrograph is only able to resolve velocity dispersions
above some limit, and (\emph{ii}) only objects brighter than some magnitude
are observed.
The selection function may then be expressed as
\be
\Pr(\calS\mid\mathscr{D},\vartheta)
\equiv W(\hx,\vartheta,\varphi)
=\Theta(\hs-s_\mrm{cut})\,\Theta(m_\mrm{cut}-\hat{m}),
\ee
where $\hat{m}$ is the observed magnitude and $\varphi$ represents selection
parameters (that may be fixed according to the instrument specifications).
This selection function $W$ is simply the statement that all objects with
$s<s_\mrm{cut}$ or $m>m_\mrm{cut}$ are not expressed  in the data.

Suppose we have observed for the $n^\mrm{th}$ object the triple
$\hat{\x}_n=(\hat{r}_n,\hat{s}_n,\hat{\imath}_n)$, with
$\hat{r}_n=\log\hat\theta_n+\log d_A(\hat{z}_n)$, where we recall
$d_A(\hat{z}_n)$ depends on the peculiar velocity $v_n$. Returning to
\eqref{eq:FP-ML} we have
\be\label{eq:fn}
f_n
= \int\dif^3\hx_n\; \calN(\hx_n\,;\, \bar{\x}, \cov^\mrm{FP}+\E_n^\mrm{FP})\,
	W(\hx_n,\vartheta,\varphi).
\ee
Clearly $f_n=1$ in the absence of selection effects. The difficulty is that
$f_n$ cannot be expressed in closed form, and the triple integral is over an
infinite domain making brute force numerical evaluation impractical. In the past
$f_n$ was estimated using expensive Monte carlo simulations
\citep{Springob:2014qja}. Here we show $f_n$ can be reduced to a two-dimensional
integral, then recast as a bivariate Gaussian probability over a rectangular
domain; this form is readily evaluated (at machine precision) using a
fast algorithm \citep{Genz:2004}.
To do this we rewrite the FP in terms of the magnitude $m$. We have that
$m$ is related to $r=\log R_e$ and $i=\log\langle I_e\rangle$ through the
average surface brightness $\langle I_e\rangle=L/(\pi R_e^2)$, where $L$ is the
luminosity. Since the absolute magnitude is $M\equiv -2.5\log L+M_0$, we have
\be
\hat{m}=-2.5(\hi+2\rhat)+\mu+M_0,
\ee
where $\mu=\hat{m}-M=5\log(d_L/\mrm{10\,pc})$ is the distance modulus, and
$d_L(\hat{z})$ is the luminosity distance. This shows that the magnitude limit
defines a diagonal cut in the space of coordinates $(\rhat,\hs,\hi)$.
If we make a change of variables to $\hat{u}\equiv\hi+2\rhat$ so
$\hx=(\rhat,\hs,\hi)\to\hat{\mbf{w}}=(\hat{u},\hs,\hi)$ we can transform this
into two orthogonal cuts in the space of coordinates $(\hat{u},\hs,\hi)$. The
integral \eqref{eq:fn} now reads
\be
f_n
=\int^\infty_{u_{\mrm{cut},n}}\dif\hat{u}_n
    \int^\infty_{s_\mrm{cut}}\dif\hs_n
    \int^\infty_{-\infty}\dif\hi_n\;
		\calN(\hat{\mbf{w}}_n\,;\,
		\mat{J}\bar{\x}, \mat{J}(\cov^\mrm{FP}+\E_n^\mrm{FP})\mat{J}^\T),
\ee
where the cutoff $u_{\mrm{cut},n}=-(m_\mrm{cut}-\mu_n+M_0)/2.5$ varies for
each galaxy due to the distance modulus; here $M_0$ is a constant
reference magnitude, $\mat{J}$ is the Jacobian and $|\det\mat{J}|=2$.
Marginalisation is now trivial for $\hi_n$ and is achieved by striking out the
corresponding rows and columns, whereupon we are left with a bivariate
Gaussian. The mean of this distribution is $(\bar{\imath}+2\bar{r},\bar{s})$
and the covariance between $\hat{w}^a$ and $\hat{w}^b$ is
\be
\big[\mat{J}(\cov^\mrm{FP}+\E_n^\mrm{FP})\mat{J}^\T\big]_{ab}
=\sum_{i,j,A=1}^3 \frac{\partial \hat{w}^a}{\partial \hat{x}^i}\,
            \frac{\partial \hat{w}^b}{\partial \hat{x}^j} \,
            \sigma^2_A\, \hat{v}^i_A\, \hat{v}^j_A,
\ee
for $a,b=1,2$. Here $\hat{v}^i_A$ is the principal axis associated with the variance
$\sigma_A^2$ of the FP (see Appendix \ref{app:geometry}).
The remaining double integral over $\hat{u}$ and $\hs$ is a (shifted)
orthant probability that is readily evaluated numerically.

One can now repeat the derivation of Section \ref{sec:bhm} with selection. Since
$f_n$ also depends on the peculiar velocity $v_n$ (among other parameters) it
cannot simply be carried through as a multiplicative constant, though the
derivation is largely the same; the only difference is that marginalisation
over $\V$ can no longer be performed analytically.
This issue will be explored further in future work, though we note here
that provided $f_n$ is not overly sensitive to $v_n$ around the
MAP estimate, then one possible work-around is to fix the peculiar velocity
dependence at $\V=\hat{\V}_\mrm{MAP}$. In that
case the overall effect under selection is to reweight the joint posterior,
i.e.\ we multiply the posterior \eqref{eq:post} by
$[\prod_n f_n]^{-1}|_{\V=\hat{\V}_\mrm{MAP}}$.

\section{Conclusions}\label{sec:conclusions}
We have presented a probabilistic framework for cosmological
inference directly from the Fundamental Plane. The main advantage of our
approach is that we are able to bypass the need for a peculiar velocity
catalogue by taking as the primary data the surface brightness, velocity
dispersion, redshift, angular size, and the angular coordinates of each
source. Because of this cosmological inference is expected to more closely
reflect the inherent uncertainties in observed data.
We emphasize that no independent distance estimate is required:
the mapping between the angular size and physical size is optimised when
performing a joint fit of the FP and the cosmological model. Our
approach thus improves upon the standard one by eliminating the need to
assume a fiducial cosmological model when converting between angular and
physical sizes during calibration.
Although it might be argued that the dependence of distance on cosmology is
weak at the low redshifts typical of peculiar velocity surveys, it should also
be noted that the peculiar velocities are not generated at random but have
a certain statistical pattern that depends very much on cosmology. These
peculiar velocities cause fluctuations in the distance and it is therefore
important to model any cosmological dependence accordingly, whether statistical
or deterministic.

We have presented a simplified cosmological analysis as a demonstration of our
method, but we emphasize that our method can be adapted to more
sophisticated analyses that have been carried out in the past, such as the
survey analyses of \cite{Johnson:2014kaa} and \cite{Howlett:2017}.
From a practical standpoint, our method is not different from the maximum
likelihood approach in how it can be used to constrain cosmology;
it can be usefully thought of as a generalisation that extends the starting
point of the analysis back to more basic inputs from which peculiar velocities
are usually estimated. No catalogue of peculiar velocities is therefore required.

Our method treats the peculiar velocities as free parameters.
For realistic data sets this will mean tens of thousands of additional
parameters, leading to an inference problem that quickly becomes intractable.
However, here peculiar velocities are only interesting in so far as what
they tell us about the underlying cosmological model. Thus in deriving the joint
posterior \eqref{eq:post} we have marginalised over them.
To do this we exploit the correlations from large-scale structure, which we
note also has the effect of regularising the inference problem.
In the linear regime that we are concerned, peculiar velocities obey Gaussian
statistics so we have used the informative prior \eqref{eq:pec-vel-like}.
The final constraints on parameters thus takes into account the considerable
uncertainties in peculiar velocities in a fully consistent way. However, as we
have shown this step can easily be omitted if one is interested in using
peculiar velocities for other purposes, and to this end we have constructed
a new MAP estimator \eqref{eq:VMAP}.

The zero-point parameter $c$ (or the degenerate parameter $\br$) is an
important parameter, as without it only relative velocities can be determined.
In the past the zero-point was calibrated by
making the assumption that the mean peculiar velocity over all galaxies is
zero \citep{Magoulas:2012jy,Springob:2014qja}. It can be seen from \eqref{eq:FP},
\eqref{eq:r}, and \eqref{eq:dA-perturbed} why such an
assumption is necessary and cannot be determined by the data: a uniform shift
$v\to v+\mrm{const}$ for all galaxies can be absorbed by the zero-point,
in exactly the same way that the degeneracy between $\HO$ and the absolute
magnitude $M$ of type Ia supernovae as distance indicators requires one to be fixed
by fiat.
In our framework, however, this assumption is redundant because the standard
prior on peculiar velocities \eqref{eq:pec-vel-like} restricts
the physically reasonable range of values of $v$, i.e.\ arbitrarily large
velocities are highly unlikely in the \LCDM\ model.

That we are able to obtain an analytic joint posterior \eqref{eq:post} is
because: (\emph{i}) peculiar velocities enter
as a linear, non-integrated fluctuation to the distance; (\emph{ii}) fitting
the FP is fitting to a linear relation; and (\emph{iii}) the underlying
source population data is well-described by a Gaussian distribution. It is
interesting to ask whether we might obtain similar analytic results when applied
to the Tully-Fisher relation since (\emph{i}) and (\emph{ii}) are satisfied
as well.
While we expect the methods described in this work (e.g.\ exploiting
correlations from large-scale structure to derive peculiar velocities) can
be applied to a Tully-Fisher-based survey, a compact Gaussian expression
for the joint posterior would seem to depend on there being some Gaussian
description of the data at hand.

In this work we have also demonstrated in Section~\ref{sec:expts} that the
primary data is sufficient to
recover cosmological parameters by generating ersatz data sets. In future
work this will be applied to more realistic data from, e.g.\ $N$-body simulations.
This will, however, require an implementation of the selection effects
described in Section~\ref{sec:selection} that restricts the observationally
accessible region of the FP.
As we saw, selection effects can be accommodated in our framework in a similar
way to how it is already dealt with using the FP maximum likelihood \eqref{eq:FP-ML}.
The posterior \eqref{eq:post} is, in effect, truncated; the normalisation is
thus modified, requiring the evaluation of a high-dimensional integral.

There are a number of other research directions we have not pursued in this work.
For example, using our joint posterior systematic biases in recovered parameters
could be investigated.
These biases occur when parameters are fixed to non-optimal values
and have non-negligible correlations with other parameters.

In future work we will present a more detailed numerical study of the
framework presented here, the performance of parameter recovery in this
approach compared with conventional analyses, a more in depth investigation
of systematic biases and its effect on building peculiar velocity catalogues,
and the application of the methods described here to real data, such as from
the upcoming Taipan survey \citep{daCunha:2017wwy}.

\section*{Acknowledgements}
The author wishes to especially thank Krzysztof Bolejko and Geraint Lewis for
helpful comments, discussion, and their support and encouragement throughout
the completion of this work. The author also thanks Matthew Colless and Nick Kaiser
for comments and feedback, and in particular Chris Blake for assistance with the
mock catalogues.
The author is supported by the Australian government Research Training Program,
and acknowledges the use of Artemis at The University of Sydney for providing HPC
resources that have contributed to the research results reported within this paper.
This work has made use of the 
publicly available codes
{\tt emcee} \citep{ForemanMackey:2012ig}\footnote{\url{https://emcee.readthedocs.io/}} and 
{\tt getdist} \citep{getdist}.\footnote{\url{https://getdist.readthedocs.io/}}

\section*{Data Availability}
The mock catalogues analysed in this work were based in part on data provided
by the CosmoSim database, a service by the Leibniz-Institute for
Astrophysics Potsdam (AIP). The MultiDark database was developed in cooperation
with the Spanish MultiDark Consolider Project CSD2009-00064.
We gratefully acknowledge the Gauss Centre for Supercomputing e.V.%
\footnote{\url{https://www.gauss-centre.eu}} and the Partnership for Advanced
Supercomputing in Europe\footnote{\url{https://www.prace-ri.eu}} for funding
the MultiDark simulation project by providing computing time on the GCS
Supercomputer SuperMUC at Leibniz Supercomputing Centre.\footnote{\url{https://www.lrz.de}}

The code used to obtain the numerical results is available at
\url{https://github.com/lhd23/BayesPV}.

\begin{multicols}{2}
\input{main.bbl}
\bibliographystyle{mnras}

\begin{thebibliography}{}
\makeatletter
\relax
\def\mn@urlcharsother{\let\do\@makeother \do\$\do\&\do\#\do\^\do\_\do\%\do\~}
\def\mn@doi{\begingroup\mn@urlcharsother \@ifnextchar [ {\mn@doi@}
  {\mn@doi@[]}}
\def\mn@doi@[#1]#2{\def\@tempa{#1}\ifx\@tempa\@empty \href
  {http://dx.doi.org/#2} {doi:#2}\else \href {http://dx.doi.org/#2} {#1}\fi
  \endgroup}
\def\mn@eprint#1#2{\mn@eprint@#1:#2::\@nil}
\def\mn@eprint@arXiv#1{\href {http://arxiv.org/abs/#1} {{\tt arXiv:#1}}}
\def\mn@eprint@dblp#1{\href {http://dblp.uni-trier.de/rec/bibtex/#1.xml}
  {dblp:#1}}
\def\mn@eprint@#1:#2:#3:#4\@nil{\def\@tempa {#1}\def\@tempb {#2}\def\@tempc
  {#3}\ifx \@tempc \@empty \let \@tempc \@tempb \let \@tempb \@tempa \fi \ifx
  \@tempb \@empty \def\@tempb {arXiv}\fi \@ifundefined
  {mn@eprint@\@tempb}{\@tempb:\@tempc}{\expandafter \expandafter \csname
  mn@eprint@\@tempb\endcsname \expandafter{\@tempc}}}

\bibitem[\protect\citeauthoryear{Abate et~al.}{2008}]{Abate:2008zy}
  Abate A., et~al., 2008, \mn@doi [\mnras] {10.1111/j.1365-2966.2008.13637.x},
  \href{https://ui.adsabs.harvard.edu/abs/2008MNRAS.389.1739A}{389, 1739}

\bibitem[\protect\citeauthoryear{Adams \& Blake}{2017}]{Adams:2017val}
  Adams C., Blake C., 2017, \mn@doi [\mnras] {{10.1093/mnras/stx1529}},
  \href{https://ui.adsabs.harvard.edu/abs/2017MNRAS.471..839A}{471, 839}

\bibitem[\protect\citeauthoryear{Alsing et~al.}{2016}]{Alsing:2016}
  Alsing J., et~al., 2016, \mn@doi [\mnras] {10.1093/mnras/stv2501},
  \href{https://ui.adsabs.harvard.edu/abs/2016MNRAS.455.4452A}{455, 4452}

\bibitem[\protect\citeauthoryear{Appleby, Shafieloo \& Johnson}{2015}]{Appleby:2014kea}
  Appleby S., Shafieloo A., Johnson A., 2015, \mn@doi [\apj] {10.1088/0004-637X/801/2/76},
  \href{https://ui.adsabs.harvard.edu/abs/2015ApJ...801...76A}{801, 76}

\bibitem[\protect\citeauthoryear{Bacon et~al.}{2014}]{Bacon:2014uja}
  Bacon D.~J., et~al., 2014, \mn@doi [\mnras] {10.1093/mnras/stu1270},
  \href{https://ui.adsabs.harvard.edu/abs/2014MNRAS.443.1900B}{443, 1900}

\bibitem[\protect\citeauthoryear{Behroozi, Wechsler \& Wu}{2013}]{Behroozi:2013}
  Behroozi P.~S., Wechsler R.~H., Wu H.-Y., 2013, \mn@doi [\apj] {10.1088/0004-637X/762/2/109},
  \href{https://ui.adsabs.harvard.edu/abs/2013ApJ...762..109B}{762, 109}

\bibitem[\protect\citeauthoryear{Bolejko et~al.}{2013}]{Bolejko:2012uj}
  Bolejko K., et~al., 2013, \mn@doi [\prl] {10.1103/PhysRevLett.110.021302},
  \href{https://ui.adsabs.harvard.edu/abs/2013PhRvL.110b1302B}{110, 021302}

\bibitem[\protect\citeauthoryear{Brewer, Foreman-Mackey \& Hogg}{2013}]{Brewer:2012gt}
  Brewer B.~J., Foreman-Mackey D., Hogg D.~W., 2013, \mn@doi [\aj] {10.1088/0004-6256/146/1/7},
  \href{https://ui.adsabs.harvard.edu/abs/2013AJ....146....7B}{146, 7}

\bibitem[\protect\citeauthoryear{Bridle et~al.}{2002}]{Bridle:2001zv}
  Bridle S.~L., et~al., 2002, \mn@doi [\mnras] {10.1046/j.1365-8711.2002.05709.x},
  \href{https://ui.adsabs.harvard.edu/abs/2002MNRAS.335.1193B}{335, 1193}

\bibitem[\protect\citeauthoryear{Campbell et~al.}{2014}]{Campbell:2014uia}
  Campbell L.~A., et~al., 2014, \mn@doi [\mnras] {10.1093/mnras/stu1198},
  \href{https://ui.adsabs.harvard.edu/abs/2014MNRAS.443.1231C}{443, 1231}

\bibitem[\protect\citeauthoryear{Carrick et~al.}{2015}]{Carrick:2015}
  Carrick J., et~al., 2015, \mn@doi [\mnras] {10.1093/mnras/stv547},
  \href{https://ui.adsabs.harvard.edu/abs/2015MNRAS.450..317C}{450, 317}

\bibitem[\protect\citeauthoryear{Colless et~al.}{2001}]{Colless:2000et}
  Colless M., et~al., 2001, \mn@doi [\mnras] {10.1046/j.1365-8711.2001.04044.x},
  \href{https://ui.adsabs.harvard.edu/abs/2001MNRAS.321..277C}{321, 277}

\bibitem[\protect\citeauthoryear{Courtois et~al.}{2013}]{Courtois:2013}
  Courtois H.~M., et~al., 2013, \mn@doi [\aj] {10.1088/0004-6256/146/3/69},
  \href{https://ui.adsabs.harvard.edu/abs/2013AJ....146...69C}{146, 69}

\bibitem[\protect\citeauthoryear{da Cunha et~al.}{2017}]{daCunha:2017wwy}
  da Cunha E., et~al., 2017, \mn@doi [Publ.\ Astron.\ Soc.\ Australia] {10.1017/pasa.2017.41},
  \href{https://ui.adsabs.harvard.edu/abs/2017PASA...34...47D}{34, 47}


\bibitem[\protect\citeauthoryear{Davis \& Scrimgeour}{2014}]{Davis:2014jwa}
  Davis T.~M., Scrimgeour M.~I., 2014, \mn@doi [\mnras] {10.1093/mnras/stu920},
  \href{https://ui.adsabs.harvard.edu/abs/2014MNRAS.442.1117D}{442, 1117}

 \bibitem[\protect\citeauthoryear{Dekel et~al.}{1999}]{Dekel:1999}
   Dekel A., et~al., 1999, ApJ, 522, 1

 \bibitem[\protect\citeauthoryear{Djorgovski \& Davis}{1987}]{Djorgovski:1987}
   Djorgovski S., Davis M., 1987, \mn@doi [\apj] {10.1086/164948},
   \href{https://ui.adsabs.harvard.edu/abs/1987ApJ...313...59D}{313, 59}

\bibitem[\protect\citeauthoryear{Dressler et~al.}{1987}]{Dressler:1987}
  Dressler A., et~al., 1987, \mn@doi [\apj] {10.1086/164947},
  \href{https://ui.adsabs.harvard.edu/abs/1987ApJ...313...42D}{313, 42}

\bibitem[\protect\citeauthoryear{Feeney, Mortlock \& Dalmasso}{2017}]{Feeney:2017sgx}
  Feeney S.~M., Mortlock D.~J., Dalmasso N., 2018, \mn@doi [\mnras] {10.1093/mnras/sty418},
  \href{https://ui.adsabs.harvard.edu/abs/2018MNRAS.476.3861F}{476, 3861}

\bibitem[\protect\citeauthoryear{Fisher et~al.}{1995}]{Fisher:1995}
  Fisher K.~B., et~al., 1995, \mn@doi [\mnras] {10.1093/mnras/272.4.885},
  \href{https://ui.adsabs.harvard.edu/abs/1995MNRAS.272..885F}{272, 885}

\bibitem[\protect\citeauthoryear{Foreman-Mackey et~al.}{2013}]{ForemanMackey:2012ig}
  Foreman-Mackey D., et~al., 2013, \mn@doi [\pasp] {10.1086/670067},
  \href{https://ui.adsabs.harvard.edu/abs/2013PASP..125..306F}{125, 306}

\bibitem[\protect\citeauthoryear{Genz}{2004}]{Genz:2004}
   Genz A., 2004, \mn@doi [Statist.\ and Comput., 14, 251] {10.1023/B:STCO.0000035304.20635.31}

\bibitem[\protect\citeauthoryear{Hellwing et~al.}{2014}]{Hellwing:2014}
  Hellwing W.~A., et~al., 2014, \mn@doi [\prl] {10.1103/PhysRevLett.112.221102},
  \href{https://ui.adsabs.harvard.edu/abs/2014PhRvL.112v1102H}{112, 221102}
   
\bibitem[\protect\citeauthoryear{Hinton et~al.}{2019}]{Hinton:2019}
  Hinton S., et~al., 2019, \mn@doi [\apj] {10.3847/1538-4357/ab13a3},
  \href{https://ui.adsabs.harvard.edu/abs/2019ApJ...876...15H}{876, 15}

\bibitem[\protect\citeauthoryear{Hoffman et~al.}{2017}]{Hoffman:2017ako}
  Hoffman Y., et~al., 2017, \mn@doi [Nature Astronomy] {10.1038/s41550-016-0036},
  \href{https://ui.adsabs.harvard.edu/abs/2017NatAs...1E..36H}{1, 0036}

\bibitem[\protect\citeauthoryear{Hogg, Myers \& Bovy}{2010}]{Hogg:2010ma}
  Hogg D.~W., Myers A.~D., Bovy J., 2010, \mn@doi [\apj] {10.1088/0004-637X/725/2/2166},
  \href{https://ui.adsabs.harvard.edu/abs/2010ApJ...725.2166H}{725, 2166}

\bibitem[\protect\citeauthoryear{Howlett et~al.}{2017}]{Howlett:2017}
  Howlett C., et~al., 2017, \mn@doi [\mnras] {10.1093/mnras/stx1521},
  \href{https://ui.adsabs.harvard.edu/abs/2017MNRAS.471.3135H}{471, 3135}

\bibitem[\protect\citeauthoryear{Hui \& Greene}{2006}]{Hui:2005nm}
  Hui L., Greene P.~B., 2006, \mn@doi [\prd] {10.1103/PhysRevD.73.123526},
  \href{https://ui.adsabs.harvard.edu/abs/2006PhRvD..73l3526H}{73, 123526}

\bibitem[\protect\citeauthoryear{Huterer et~al.}{2017}]{Huterer:2016uyq}
  Huterer D., et~al., 2017, \mn@doi [\jcap] {10.1088/1475-7516/2017/05/015},
  \href{https://ui.adsabs.harvard.edu/abs/2017JCAP...05..015H}{05, 015}

\bibitem[\protect\citeauthoryear{Jasche et~al.}{2010}]{Jasche:2010}
  Jasche J., et~al., 2010, \mn@doi [\mnras] {10.1111/j.1365-2966.2010.16610.x},
  \href{https://ui.adsabs.harvard.edu/abs/2010MNRAS.406...60J}{406, 60}

\bibitem[\protect\citeauthoryear{Jasche \& Wandelt}{2013}]{Jasche:2013}
  Jasche J., Wandelt B.~D., 2013, \mn@doi [\apj] {10.1088/0004-637X/779/1/15},
  \href{https://ui.adsabs.harvard.edu/abs/2013ApJ...779...15J}{779, 15}
  
\bibitem[\protect\citeauthoryear{Jaffe \& Kaiser}{1995}]{Jaffe:1994gx}
  Jaffe A.~H., Kaiser N., 1995, \mn@doi [\apj] {10.1086/176551},
  \href{https://ui.adsabs.harvard.edu/abs/1995ApJ...455...26J}{455, 26}

\bibitem[\protect\citeauthoryear{Johnson et~al.}{2014}]{Johnson:2014kaa}
  Johnson A., et~al., 2014, \mn@doi [\mnras] {10.1093/mnras/stw447},
  \href{https://ui.adsabs.harvard.edu/abs/2016MNRAS.458.2725J}{444, 3926}

\bibitem[\protect\citeauthoryear{Johnson et~al.}{2016}]{Johnson:2015aaa}
  Johnson A., et~al., 2016, \mn@doi [\mnras] {10.1093/mnras/stu1615},
  \href{https://ui.adsabs.harvard.edu/abs/2014MNRAS.444.3926J}{458, 2725}

\bibitem[\protect\citeauthoryear{Kaiser \& Hudson}{2015}]{Kaiser:2014jca}
  Kaiser N., Hudson M.~J., 2015, \mn@doi [\mnras] {10.1093/mnras/stv693},
  \href{https://ui.adsabs.harvard.edu/abs/2015MNRAS.450..883K}{450, 883}

\bibitem[\protect\citeauthoryear{Klypin et al.}{2016}]{Klypin:2016}
  Klypin A., et~al., 2016, \mn@doi [\mnras] {10.1093/mnras/stw248},
  \href{https://ui.adsabs.harvard.edu/abs/2016MNRAS.457.4340K}{457, 4340}

\bibitem[\protect\citeauthoryear{Koda et~al.}{2014}]{Koda:2013eya}
  Koda J., et~al., 2014, \mn@doi [\mnras] {10.1093/mnras/stu1610},
  \href{https://ui.adsabs.harvard.edu/abs/2014MNRAS.445.4267K}{445, 4267}

\bibitem[\protect\citeauthoryear{Leistedt, Mortlock \& Peiris}{2016}]{Leistedt:2016}
  Leistedt B., Mortlock D.~J., Peiris H.~V., 2016, \mn@doi [\mnras] {10.1093/mnras/stw1304},
  \href{https://ui.adsabs.harvard.edu/abs/2016MNRAS.460.4258L}{460, 4258}

\bibitem[\protect\citeauthoryear{Lewis}{2019}]{getdist}
   Lewis A., 2019, preprint (\mn@eprint{arXiv}{1910.13970})

\bibitem[\protect\citeauthoryear{Loredo}{2004}]{Loredo:2004}
  Loredo T.~J., 2004, \mn@doi [AIPC] {10.1063/1.1835214},
  \href{https://ui.adsabs.harvard.edu/abs/2004AIPC..735..195L}{735, 195}

\bibitem[\protect\citeauthoryear{Loredo \& Hendry}{2010}]{Hobson:2010}
   Loredo T.~J., Hendry M.~A., 2010, in Hobson M.~P.\ et~al., eds., 
   Bayesian Methods in Cosmology, Cambridge University Press, Cambridge, p.~245

\bibitem[\protect\citeauthoryear{Loredo}{2012}]{Loredo:2012}
   Loredo T.~J., 2012, in Hilbe J.~M., ed., 
   Astrostatistical Challenges for the New Astronomy,
   Springer, New York

\bibitem[\protect\citeauthoryear{Ma, Gordon \& Feldman}{2011}]{Ma:2010ps}
  Ma Y.~Z., Gordon C., Feldman H.~A., 2011, \mn@doi [\prd] {10.1103/PhysRevD.83.103002},
  \href{https://ui.adsabs.harvard.edu/abs/2011PhRvD..83j3002M}{83, 103002}

\bibitem[\protect\citeauthoryear{Macauley et~al.}{2012}]{Macaulay:2011av}
  Macaulay E., et~al., 2012, \mn@doi [\mnras] {10.1111/j.1365-2966.2012.21629.x},
  \href{https://ui.adsabs.harvard.edu/abs/2012MNRAS.425.1709M}{425, 1709}

\bibitem[\protect\citeauthoryear{Mandel et~al.}{2009}]{Mandel:2009}
  Mandel K.~S., et~al., 2009, \mn@doi [\apj] {10.1088/0004-637X/704/1/629},
  \href{https://ui.adsabs.harvard.edu/abs/2009ApJ...704..629M}{704, 629}

\bibitem[\protect\citeauthoryear{March et~al.}{2011}]{March:2011}
  March M.~C.et~al., 2011, \mn@doi [\mnras] {10.1111/j.1365-2966.2011.19584.x},
  \href{https://ui.adsabs.harvard.edu/abs/2011MNRAS.418.2308M}{418, 230}

\bibitem[\protect\citeauthoryear{Magoulas et~al.}{2012}]{Magoulas:2012jy}
  Magoulas C., et~al., 2012, \mn@doi [\mnras] {10.1111/j.1365-2966.2012.21421.x},
  \href{https://ui.adsabs.harvard.edu/abs/2012MNRAS.427..245M}{427, 245}

\bibitem[\protect\citeauthoryear{Mandel, Farr \& Gair}{2019}]{Mandel:2018mve}
  Mandel I., Farr W.~M., Gair J.~R., 2019, \mn@doi [\mnras] {10.1093/mnras/stz896},
  \href{https://ui.adsabs.harvard.edu/abs/2019MNRAS.486.1086M}{486, 1086}

\bibitem[\protect\citeauthoryear{Nusser \& Davis}{2011}]{Nusser:2011tu}
  Nusser A., Davis M., 2011, \mn@doi [\apj] {10.1088/0004-637X/736/2/93},
  \href{https://ui.adsabs.harvard.edu/abs/2011ApJ...736...93N}{736, 93}

\bibitem[\protect\citeauthoryear{Nusser}{2017}]{Nusser:2017}
  Nusser A., 2017, \mn@doi [\mnras] {10.1093/mnras/stx1225},
  \href{https://ui.adsabs.harvard.edu/abs/2017MNRAS.470..445N}{470, 445}

\bibitem[\protect\citeauthoryear{Okumura et al.}{2014}]{Okumura:2014}
  Okumura T., et~al., 2014, \mn@doi [\jcap] {10.1088/1475-7516/2014/05/003}
  \href{https://ui.adsabs.harvard.edu/abs/2014JCAP...05..003O}{2014, 003}

\bibitem[\protect\citeauthoryear{Peebles}{1993}]{Peebles:1993}
  Peebles P.~J.~E., 1993, Principles of Physical Cosmology, Princeton Univ.~Press, Princeton, NJ

\bibitem[\protect\citeauthoryear{Petersen \& Pedersen}{2006}]{Peterson:2006}
   Petersen K.~B., Pedersen M.~S., 2006, The matrix cookbook

\bibitem[\protect\citeauthoryear{Planck Collaboration et~al.}{2015}]{Likelihood:2015}
  Planck Collaboration, et~al., 2016,  \mn@doi [\aap] {10.1051/0004-6361/201526926},
  \href{https://ui.adsabs.harvard.edu/abs/2016A&A...594A..11P}{594, A11}

\bibitem[\protect\citeauthoryear{Portillo et~al.}{2017}]{Portillo:2017}
  Portillo S.~K.~N., Lee B.~C.~G., Daylan T., Finkbeiner D.~P., 2017, \mn@doi [\aj] {10.3847/1538-3881/aa8565},
  \href{https://ui.adsabs.harvard.edu/abs/2017AJ....154..132P}{154, 132}

\bibitem[\protect\citeauthoryear{Qin, Howlett \& Staveley-Smith}{2019}]{Qin:2019}
  Qin F., Howlett C., Staveley-Smith L., 2019, \mn@doi [\mnras] {10.1093/mnras/stz1576},
  \href{https://ui.adsabs.harvard.edu/abs/2019MNRAS.487.5235Q}{487, 5235}

\bibitem[\protect\citeauthoryear{Saglia et~al.}{2001}]{Saglia:2000fa}
  Saglia R.~P., et~al., 2001, \mn@doi [\mnras] {10.1046/j.1365-8711.2001.04317.x},
  \href{https://ui.adsabs.harvard.edu/abs/2001MNRAS.324..389S}{324, 389}

\bibitem[\protect\citeauthoryear{S{\'a}nchez \& Bernstein}{2019}]{Sanchez:2019}
  S{\'a}nchez C., Bernstein G.~M., 2019, \mn@doi [\mnras] {10.1093/mnras/sty3222},
  \href{https://ui.adsabs.harvard.edu/abs/2019MNRAS.483.2801S}{483, 2801}

\bibitem[\protect\citeauthoryear{Sasaki}{1987}]{Sasaki:1987ad}
  Sasaki M., 1987, \mn@doi [\mnras] {10.1093/mnras/228.3.653},
  \href{https://ui.adsabs.harvard.edu/abs/1987MNRAS.228..653S}{228, 653}

\bibitem[\protect\citeauthoryear{Schwarz \& Weinhorst}{2007}]{Schwarz:2007wf}
  Schwarz D.~J., Weinhorst B., 2007, \mn@doi [\aap] {10.1051/0004-6361:20077998},
  \href{https://ui.adsabs.harvard.edu/abs/2007A&A...474..717S}{474, 717}

\bibitem[\protect\citeauthoryear{Schneider et~al.}{2015}]{Schneider:2015}
  Schneider M.~D., et~al., 2015, \mn@doi [\apj] {10.1088/0004-637X/807/1/87},
  \href{https://ui.adsabs.harvard.edu/abs/2015ApJ...807...87S}{807, 87}

\bibitem[\protect\citeauthoryear{Scrimgeour et~al.}{2016}]{Scrimgeour:2015khj}
  Scrimgeour M.~I., et~al., 2016, \mn@doi [\mnras] {10.1093/mnras/stv2146},
  \href{https://ui.adsabs.harvard.edu/abs/2016MNRAS.455..386S}{455, 386}

\bibitem[\protect\citeauthoryear{Sharif et~al.}{2016}]{Sharif:2016}
  Sharif H., et~al., 2016, \mn@doi [\apj] {10.3847/0004-637X/827/1/1},
  \href{https://ui.adsabs.harvard.edu/abs/2016ApJ...827....1S}{827, 1}

\bibitem[\protect\citeauthoryear{Soltis et~al.}{2019}]{Soltis:2019ryf}
  Soltis J., et~al., 2019, \mn@doi [\prl] {10.1103/PhysRevLett.122.091301},
  \href{https://ui.adsabs.harvard.edu/abs/2019PhRvL.122i1301S}{122, 091303}

\bibitem[\protect\citeauthoryear{Springel}{2005}]{Springel:2005}
  Springel V., 2005, \mn@doi [\mnras] {10.1111/j.1365-2966.2005.09655.x},
  \href{https://ui.adsabs.harvard.edu/abs/2005MNRAS.364.1105S}{364, 1105}
  
\bibitem[\protect\citeauthoryear{Springob et~al.}{2014}]{Springob:2014qja}
  Springob C.~M., et~al., 2014, \mn@doi [\mnras] {10.1093/mnras/stu1743},
  \href{https://ui.adsabs.harvard.edu/abs/2014MNRAS.445.2677S}{445, 2677}

\bibitem[\protect\citeauthoryear{Strauss \& Willick}{1995}]{Strauss:1995fz}
  Strauss M.~A, Willick J.~A., 1995, \mn@doi [\physrep] {10.1016/0370-1573(95)00013-7},
  \href{https://ui.adsabs.harvard.edu/abs/1995PhR...261..271S}{261, 271}

\bibitem[\protect\citeauthoryear{Sugiura, Sugiyama \& Sasaki}{1999}]{Sugiura:1999a} 
  Sugiura N., Sugiyama N., Sasaki M., 1999, \mn@doi [Progress Theor.\ Phys.] {10.1143/PTP.101.903},
  \href{https://ui.adsabs.harvard.edu/abs/1999PThPh.101..903S}{101, 903}

\bibitem[\protect\citeauthoryear{Tully, Courtois \& Sorce}{2016}]{Tully:2016ppz}
  Tully R.~B., Courtois H.~M., Sorce J.~G., 2016, \mn@doi [\aj] {10.3847/0004-6256/152/2/50},
  \href{https://ui.adsabs.harvard.edu/abs/2016AJ....152...50T}{152, 50}

\bibitem[\protect\citeauthoryear{Willick}{1994}]{Willick:1994}
  Willick J.~A., 1994, \mn@doi [\apjs] {10.1086/191957},
  \href{https://ui.adsabs.harvard.edu/abs/1994ApJS...92....1W}{92, 1}

\bibitem[\protect\citeauthoryear{Willick et~al.}{1997}]{Willick:1996km}
  Willick J.~A., et~al., 1997, \mn@doi [\apj] {10.1086/304551},
  \href{https://ui.adsabs.harvard.edu/abs/1997ApJ...486..629W}{486, 629}

\bibitem[\protect\citeauthoryear{Willick \& Strauss}{1998}]{Willick:1998}
  Willick J.~A., Strauss M.~A., 1998, \mn@doi [\apj] {10.1086/306314},
  \href{https://ui.adsabs.harvard.edu/abs/1998ApJ...507...64W}{507, 64}

\bibitem[\protect\citeauthoryear{Zaroubi, Hoffman \& Dekel}{1999}]{Zaroubi:1999}
  Zaroubi S., Hoffman Y., Dekel A., 1999, \mn@doi [\apj] {10.1086/307473},
  \href{https://ui.adsabs.harvard.edu/abs/1999ApJ...520..413Z}{520, 413}

\bibitem[\protect\citeauthoryear{Zhang et~al.}{2017}]{Zhang:2017}
  Zhang B.~R., et~al., 2017, \mn@doi [\mnras] {10.1093/mnras/stx1600},
  \href{https://ui.adsabs.harvard.edu/abs/2017MNRAS.471.2254Z}{471, 2254}

\makeatother
\end{thebibliography}
\end{multicols}

\appendix

\section{Geometry of the Fundamental Plane}\label{app:geometry}

The trivariate Gaussian in the space $(r,s,i)$ can be thought of as
a 3-dimensional ellipsoid, with principal axes given by
\begin{subequations}
\begin{align}
\hat{\mbf{v}}_1 &= 1/\sqrt{1+a^2+b^2}\,(1, -a, -b)^\T, \\[3pt]
\hat{\mbf{v}}_2 &= 1/\sqrt{1+b^2}\,(b, 0, 1)^\T, \\[3pt]
\hat{\mbf{v}}_3 &= 1/\sqrt{(1+b^2)(1+a^2+b^2)}\,(-a, -1-b^2, ab)^\T.
\end{align}
\end{subequations}
Here $\hat{\mbf{v}}_1$ is normal to the FP, and $\hat{\mbf{v}}_2$ and $\hat{\mbf{v}}_3$
span it. Note $\{\hat{\mbf{v}}_1,\hat{\mbf{v}}_2,\hat{\mbf{v}}_3\}$ form an
orthonormal basis. From \eqref{eq:FP} $\hat{\mbf{v}}_1$ is given (up to an
overall sign change)
but only determines $\hat{\mbf{v}}_2$ and
$\hat{\mbf{v}}_3$ up to a rotation about $\hat{\mbf{v}}_1$. Following
\cite{Magoulas:2012jy} we have chosen $\hat{\mbf{v}}_2$ so that it has
vanishing $s$ component. Note we do not make any assumptions about
$\hat{\mbf{v}}_2$ or $\hat{\mbf{v}}_3$ when we fit the FP; there are then nine free
parameters, three for the mean, and six for the covariances.
All 3D vectors are specified in the order
$(r,s,i)\in\mathbb{R}^3$.

Define $\x=(r,s,i)^\T$ and $\mbf{u}=\mat{O}^\T\x$
where
\be
\mat{O}=
(\hat{\mbf{v}}_1,\hat{\mbf{v}}_2,\hat{\mbf{v}}_3)=
\begin{pmatrix}
v_{1,1} & v_{2,1} &v_{3,1} \\
v_{1,2} & v_{2,2} &v_{3,2} \\
v_{1,3} & v_{2,3} &v_{3,3}
\end{pmatrix}
\ee
is an orthogonal matrix ($\mat{O}\mat{O}^\T=\mat{O}^\T\mat{O}=\mat{I}$).
The covariance matrix of $\x$ is
\begin{align}
\cov
&=\big\langle(\x-\xbar)(\x-\xbar)^\T\big\rangle \nonumber\\[3pt]
&=\mat{O}\big\langle(\mbf{u}-\bar{\mbf u})
	(\mbf{u}-\bar{\mbf u})^\T\big\rangle\mat{O}^\T
\equiv\mat{O}\mat{D}\mat{O}^\T,
\label{eq:C}
\end{align}
where $\mat{D}=\mathrm{diag}(\sigma_1^2,\sigma_2^2,\sigma_3^2)$.
The components relative to the principal axes can be computed by projection:
\begin{subequations}
\begin{align}
\mathbf{u}
&=\mat{O}^\T\x
=(\hat{\mbf{v}}_1\cdot\x,\, \hat{\mbf{v}}_2\cdot\x,\,
	\hat{\mbf{v}}_3\cdot\x)^\T, \\[3pt]
\bar{\mathbf u}
&=\mat{O}^\T\xbar
=(\hat{\mbf{v}}_1\cdot\xbar,\, \hat{\mbf{v}}_2\cdot\xbar,\,
	\hat{\mbf{v}}_3\cdot\xbar)^\T.
\end{align}
\end{subequations}

\section{Review of some properties of Gaussians}
Here we list some useful formulae involving Gaussians relevant to our calculations.
These standard results can be found in, e.g.\ \cite{Peterson:2006}.%

Let $\x$ be an $N$-dimensional random vector. An $N$-dimensional multivariate
Gaussian density with mean $\bm\mu$ and covariance $\Sig$ shall be denoted
\be
\calN(\x\, ;\, \bm\mu,\Sig)
\equiv(2\pi)^{-N/2}\det\Sig^{-1/2}
	\exp\left[-\frac{1}{2}(\x-\bm\mu)^\T\Sig^{-1}(\x-\bm\mu)\right].
\ee

The product of two multivariate Gaussians gives a scaled Gaussian
\be\label{eq:gauss-prod}
\calN(\x\, ;\, \bm{\mu}_1, \Sig_1)\, \calN(\x\,;\, \bm{\mu}_2, \Sig_2)
	= Z \, \calN(\x\, ;\, \bm{\mu}_3, \Sig_3),
\ee
where
\begin{align*}
&\Sig_3^{-1} =  \Sig_1^{-1} + \Sig_2^{-1}, \\
&\bm{\mu}_3 = \Sig_3\, (\Sig_1^{-1}\bm{\mu}_1 + \Sig_2^{-1}\bm{\mu}_2),
\end{align*}
and
\be
Z = (2\pi)^{-N/2}\det(\Sig_1+\Sig_2)^{-1/2}
\exp\left[-\frac{1}{2}(\bm\mu_1-\bm\mu_2)^\T(\Sig_1+\Sig_2)^{-1}(\bm\mu_1-\bm\mu_2)\right].
\ee
Observe that $\Sig_3$ is given by the harmonic sum of $\Sig_1$ and $\Sig_2$,
and that $\bm\mu_3$ is given by a weighted average of $\bm\mu_1$
and $\bm\mu_2$. The integral over $\x$ follows immediately from
\eqref{eq:gauss-prod}:
\begin{align}
\int\dif\x\:
	 \calN(\x\, ;\, \bm{\mu}_1,\Sig_1) \, \calN(\mbf{x}\, ;\, \bm{\mu}_2, \Sig_2)
&= Z\, \int\dif\x\:\calN(\x\, ;\, \bm{\mu}_3, \Sig_3) \nonumber\\[4pt]
&= \calN(\bm{\mu}_1\, ;\, \bm{\mu}_2, \, \Sig_1+\Sig_2) 
= \calN(\bm{\mu}_2\, ;\, \bm{\mu}_1, \, \Sig_1+\Sig_2)
	\label{eq:gauss-conv}
\end{align}
i.e.\ a constant of Gaussian form.
As we will frequently encounter integrals of this form we have written the
constant $Z$ using notation for a Gaussian distribution; it should not, however,
be understood as a probability density function of $\bm\mu_1$ nor $\bm\mu_2$.

The normalisation of the Gaussian,
\be
\int \dif\x\,(2\pi)^{-N/2}\det\Sig^{-1/2}\,
\exp\left[-\frac{1}{2}(\x-\bm\mu)^\T\,\Sig^{-1}\,(\x-\bm\mu)\right]
=1,
\ee
implies two other useful integrals
\begin{align}
&\int\dif\x\,\exp\left(-\frac{1}{2}\x^\T\,\Sig^{-1}\,\x+\bm\mu^\T\,\Sig^{-1}\,\x\right)
=(2\pi)^{N/2}\det\Sig^{1/2}\exp\left(\frac{1}{2}\bm\mu^\T\,\Sig^{-1}\,\bm\mu\right),
\label{eq:gauss-int1}\\[4pt]
&\int\dif\x\,\exp\left(-\frac{1}{2}\x^\T\,\A\,\x+\mbf{b}^\T\,\x\right)
=(2\pi)^{N/2}\det(\A^{-1})^{1/2}\exp\left(\frac{1}{2}\mbf{b}^\T\,\A^{-1}\,\mbf{b}\right).
\label{eq:gauss-int2}
\end{align}

The following identities related to matrix inverses are useful in
simplifying expressions.
If $\A$ and $\mat{B}$ are nonsingular
\be\label{eq:Woodbury}
(\A+\mat{U}\mat{B}\mat{V})^{-1}
=\A^{-1}
- \A^{-1}\mat{U}(\mat{B}^{-1}+\mat{V}\A^{-1}\mat{U})^{-1}\mat{V}\A^{-1}.
\ee
This is the Woodbury identity.
Another useful identity is a variant of this: If $\A+\mat{B}$ is
nonsingular then
\be\label{eq:Wood-2}
\A-\A(\A+\mat{B})^{-1} \A
=\mat{B}-\mat{B}(\A+\mat{B})^{-1} \mat{B}.
\ee
We will also make use of
\be\label{eq:inv-id2}
(\A+\mat{B})^{-1}
=\A^{-1}(\A^{-1}+\mat{B}^{-1})^{-1}\mat{B}^{-1}
=\mat{B}^{-1}(\A^{-1}+\mat{B}^{-1})^{-1}\A^{-1}.
\ee

\subsection{Conditional Gaussians\label{app:cond-gauss}}
Let $\w$ be a random vector partitioned as
\be
\w =
\begin{pmatrix}
\w_1 \\ \w_2
\end{pmatrix}
\ee
and
\be
\bm{\mu}_1=\langle \w_1\rangle=0,\qquad
\bm{\mu}_2=\langle \w_2\rangle=0,\qquad
\bm\mu=\langle\w\rangle = \begin{pmatrix} \bm\mu_1 \\ \bm\mu_2 \end{pmatrix},
\ee
be the mean vectors. The covariance is given by
\be
\Sig\equiv\langle(\w-\langle \w\rangle)(\w-\langle \w\rangle)^\T\rangle
=\langle \w\w^\T\rangle=
\begin{pmatrix}
\Sig_{11} & \Sig_{12} \\
\Sig_{21} & \Sig_{22}
\end{pmatrix},
\ee
where $\Sig_{11}=\langle\w_1\w_1^\T\rangle$, 
$\Sig_{12}=\langle\w_1\w_2^\T\rangle$, and $\Sig_{12}^\T=\Sig_{21}$.

A standard result of the statistics of multivariate Gaussians is that the
probability of $\w_1$ conditioned on $\w_2$ is given by a Gaussian
with mean
\be
\bm{\mu}_{1|2} = \bm{\mu}_1-\Sig_{12}\Sig_{22}^{-1}(\bm{\mu}_2-\w_2)
\ee
and covariance
\be
\Sig_{11|2} 
= \Sig_{11} - \Sig_{12}\Sig_{22}^{-1}\Sig_{21}.
\ee

\section{Direct calculation of the joint posterior}\label{app:details}
In this section we present details on obtaining \eqref{eq:post}. This
comes down to performing the integral
\be
\Pr(\calC\mid\zhat,\thhat,\shat,\ihat)
\propto
\Pr(\calC)
\int
    \dif\y\:
	\calN\big(\y \mid \ybar, \Cyy \big) \cdot
	\calN\big(\y \mid \yhat, \Eyy\big) \cdot
	\calN\big(\bm\Delta_r\mid \bm{0}, \Sigrr\big),
\ee
where we recall that $\bm\Delta_r$ is given by \eqref{eq:Delta-r} and depends on
$\y$, and $\Sigrr$ is the shifted theoretical covariance from both
LSS and the FP relation, given by \eqref{eq:Sig-rr}.
Here the integrand is given by the product of two $2N$-dimensional Gaussians
and one $N$-dimensional Gaussian; except for $\Pr(\calC)$, all terms depend on
$\y$. The integration is perhaps most easily performed if we rewrite the
integrand in canonical form so that
\be
\Pr(\calC\mid\zhat,\thhat,\shat,\ihat)
\propto
	\,\Pr(\calC)\,
	\det\mathbf{\Sigma}_{\mathbf{rr}}^{-1/2}\,
	\det\mathbf{C}_{\mathbf{yy}}^{-1/2}\,
	\det\mathbf{E}_{\mathbf{yy}}^{-1/2}
\int\dif\y\,
	\left[K\exp\left(\bm\eta^\T\,\y-\frac{1}{2}\y^\T\,\bm\Lambda\,\y\right)\right],
\ee
where
\be\nonumber
K\exp\left(\bm\eta^\T\,\y-\frac{1}{2}\y^\T\,\bm\Lambda\,\y\right) 
\equiv
\exp\left[
    -\frac{1}{2}(\y-\ybar)^\T\,\Cyyi\,(\y-\ybar)
    -\frac{1}{2}(\y-\yhat)^\T\,\Eyyi\,(\y-\yhat)
    -\frac{1}{2}\bm\Delta_r^\T\,\Sigrri\,\bm\Delta_r \right]
\ee
Expanding all forms on the right hand side then rearranging, we can
make the identifications
\begin{subequations}
\begin{align}
\bm\Lambda
&\equiv
	\Cyyi+\Eyyi
	+\Cyyi\,\Cry\,\Sigrri\,\CryT\,\Cyyi, \\[9pt]
\bm\eta
&\equiv
	\Cyyi\,\ybar + \Eyyi\,\yhat
	+\Cyyi\,\Cry\,\Sigrri\,(\wbar-\DelO), \\[4pt]
K&\equiv
\exp\left[
	-\frac{1}{2}\ybar^\T\,\Cyyi\,\ybar
	-\frac{1}{2}\yhat^\T\,\Eyyi\,\yhat
	-\frac{1}{2}(\wbar-\DelO)^\T\,\Sigrri\,(\wbar-\DelO)\right],
\end{align}
\end{subequations}
where we have introduced the shorthands
\begin{subequations}
\begin{align}
\DelO&\equiv
	\rbar - \big(\bm{\logl_\theta} + \bm{\logl_{\bar{d}}}\big), \\[9pt]
\wbar&\equiv
	\CryT\,\Cyyi\,\ybar,
\end{align}
\end{subequations}
so that $\bm\Delta_r=\CryT\Cyyi\y-(\wbar-\DelO)$.
The integral is now in a standard form that is readily evaluated
using \eqref{eq:gauss-int2}:
\be
\int\dif\y\,
	\exp\left(\bm\eta^\T\,\y-\frac{1}{2}\y^\T\,\bm\Lambda\,\y\right)
=(2\pi)^N\det\bm\Lambda^{1/2}
	\exp\left(\frac{1}{2}\bm\eta^\T\,\bm\Lambda^{-1}\,\bm\eta\right).
\ee
In what follows it will be convenient to introduce the shorthands
\begin{subequations}
\begin{align}
&\M\equiv \Cyyi\,\Cry\,\Sigrri\,\CryT\,\Cyyi, \\[4pt]
&\Sigyy\equiv (\Cyyi+\Eyyi)^{-1}, \\[4pt]
&\bm\mu \equiv \Sigyy\,(\Cyyi\,\ybar+\Eyyi\,\yhat), \\[4pt]
&\mutilde \equiv \CryT \Cyyi \bm\mu, \\[4pt]
&\Sig\equiv \Sigrr+\CryT\Cyyi\Sigyy\Cyyi\Cry. \label{eq:Sig}
\end{align}
\end{subequations}
We can then write
\begin{subequations}
\begin{align}
\bm\Lambda &= \Sigyyi + \M, \label{eq:Lambda}\\[8pt]
\bm\eta
&= \Sigyyi\bm\mu + \Cyyi\Cry\Sigrri(\wbar-\DelO).
\end{align}
\end{subequations}

Using \eqref{eq:inv-id2} it can be shown that
\be
\frac{1}{2}\ybar^\T\,\Cyyi\,\ybar + \frac{1}{2}\yhat^\T\,\Eyyi\,\yhat
=\frac{1}{2}\bm\mu^\T\,\Sigyyi\,\bm\mu
    +\frac{1}{2}(\yhat-\ybar)^\T\,(\Cyy+\Eyy)^{-1}\,(\yhat-\ybar)
\ee
so
\be
K=
\exp\left[
	-\frac{1}{2}\bm\mu^\T\,\Sigyyi\,\bm\mu
	-\frac{1}{2}(\wbar-\DelO)^\T\,\Sigrri\,(\wbar-\DelO)
	\right]
\exp\left[
	-\frac{1}{2}(\yhat-\ybar)^\T\,(\Cyy+\Eyy)^{-1}\,(\yhat-\ybar)
	\right].
\ee
Observe the second term of $K$ is the Gaussian \eqref{eq:p-si}
convolved with the error distribution.

The joint posterior now reads
\begin{align}
\Pr(\calC\mid\zhat,\thhat,\shat,\ihat)
&\propto
\Pr(\calC)\,
\det(\Cyy+\Eyy)^{-1/2}
\exp\left[
	-\frac{1}{2}(\yhat-\ybar)^\T\,(\Cyy+\Eyy)^{-1}\,(\yhat-\ybar)
	\right] \nonumber\\[4pt]
&\times
\big(\det\Sigyy\,\det\Sigrr\,\det\bm\Lambda\big)^{-1/2}
\exp\left[
	-\frac{1}{2}\bm\mu^\T\,\Sigyyi\,\bm\mu
	-\frac{1}{2}(\DelO-\wbar)^\T\,\Sigrri\,(\DelO-\wbar)
	+\frac{1}{2}\bm\eta^\T\,\bm\Lambda^{-1}\,\bm\eta
	\right]
	\label{eq:p-post}
\end{align}

Using the matrix determinant lemma it can be shown that
$\det\Sigyy\,\det\Sigrr\,\det\bm\Lambda=\det\Sig$.
With the Woodbury identity \eqref{eq:Woodbury} the inverse of
$\bm\Lambda$ can be written as
\be
\bm\Lambda^{-1}
=\Sigyy - \Sigyy\,\Cyyi\,\Cry\,\Sig^{-1}\,\CryT\,\Cyyi\,\Sigyy
\ee
and the inverse of $\Sig$ as
\be\label{eq:Sig-inv}
\Sig^{-1}
=\Sigrri-\Sigrri\,\CryT\,\Cyyi\,\bm\Lambda^{-1}\,\Cyyi\,\Cry\,\Sigrri.
\ee
Using \eqref{eq:Wood-2} the last expression can be rearranged to
give the useful identity
\be\label{eq:inv-id}
\Sigyyi - \Sigyyi \, \bm\Lambda^{-1} \, \Sigyyi
= \M - \M\,\bm\Lambda^{-1}\,\M
= \Cyyi\, \Cry\, \Sig^{-1}\, \CryT\, \Cyyi.
\ee
After a lengthy, but straightforward, calculation using the
equations above we find that \eqref{eq:p-post} simplifies to
\be
\Pr(\calC\mid\zhat,\thhat,\shat,\ihat)
\propto
\det\Sig^{-1/2}\,
\det(\Cyy+\Eyy)^{-1/2}\,
\exp\left[
    -\frac{1}{2}\bm\Delta^\T\,\Sig^{-1}\,\bm\Delta
	-\frac{1}{2}(\yhat-\ybar)^\T\,(\Cyy+\Eyy)^{-1}\,(\yhat-\ybar)
	\right]\,\Pr(\calC).
\ee
Because the $2N\times 2N$ matrix $\Cyy+\Eyy$ has the structure that, when
partitioned into four $N\times N$ blocks, each block is a
diagonal matrix, we can rearrange the rows and columns of the second
quadratic form to obtain the final form \eqref{eq:post}.

\bsp	
\label{lastpage}
\end{document}